\documentclass[10pt,twocolumn,preprintnumbers,amsmath,amssymb,nofootinbib
,superscriptaddress]{revtex4-1}
\usepackage{graphicx,longtable,mathrsfs,color,array}
\usepackage[hidelinks]{hyperref}
\usepackage[usenames,dvipsnames]{xcolor} 
\usepackage{amssymb,amsmath,mathtools,mathrsfs,slashed} 
\usepackage{epsfig,subfigure,placeins,float} 
\usepackage{booktabs,longtable,ctable,multirow} 
\usepackage{exscale,relsize} 
\usepackage[normalem]{ulem} 
\usepackage{enumerate}
\usepackage{times, mathptmx} 
\usepackage[utf8]{inputenc}
\usepackage{color}
\usepackage{graphicx}
\usepackage{color}
\usepackage{graphicx,graphics}
\usepackage{bm}
\usepackage{tabularx}
\usepackage{rotating}
\usepackage{units}
\allowdisplaybreaks[1]
\usepackage{hhline}
\usepackage{longtable}
\usepackage{placeins}
\usepackage{natbib}
\begin{document}

\title[Dust mass in SN 2012aw and iPTF14hls]{Quantifying the dust in SN 2012aw and iPTF14hls with ORBYTS}

\author{Maria Niculescu-Duvaz}
\affiliation{Dept. of Physics \& Astronomy, University College London, Gower St, London, WC1E 6BT,UK}
\author{M. J. Barlow}
\affiliation{Dept. of Physics \& Astronomy, University College London, Gower St, London, WC1E 6BT,UK}
\author{W. Dunn}
\affiliation{Mullard Space Science Laboratory, University College London, Holmbury St. Mary, Dorking, Surrey, RH5 6NT, UK}
\author{A. Bevan}
\affiliation{Dept. of Physics \& Astronomy, University College London, Gower St, London, WC1E 6BT,UK}
\author{Omar Ahmed}
\affiliation{Highams Park School, London, E4 9PJ,UK}
\author{David Arkless}
\affiliation{Highams Park School, London, E4 9PJ,UK}
\author{Jon Barker}
\affiliation{Highams Park School, London, E4 9PJ,UK}
\author{Sidney Bartolotta}
\affiliation{Highams Park School, London, E4 9PJ,UK}
\author{Liam Brockway}
\affiliation{Highams Park School, London, E4 9PJ,UK}
\author{Daniel Browne}
\affiliation{Highams Park School, London, E4 9PJ,UK}
\author{Ubaid Esmail}
\affiliation{Highams Park School, London, E4 9PJ,UK}
\author{Max Garner}
\affiliation{Highams Park School, London, E4 9PJ,UK}
\author{Wiktoria Guz}
\affiliation{Highams Park School, London, E4 9PJ,UK}
\author{Scarlett King}
\affiliation{Highams Park School, London, E4 9PJ,UK}
\author{Hayri Kose}
\affiliation{Highams Park School, London, E4 9PJ,UK}
\author{Madeline Lampstaes-Capes}
\affiliation{Highams Park School, London, E4 9PJ,UK}
\author{Joseph Magen}
\affiliation{Highams Park School, London, E4 9PJ,UK}
\author{Nicole Morrison}
\affiliation{Highams Park School, London, E4 9PJ,UK}
\author{Kyaw Oo}
\affiliation{Highams Park School, London, E4 9PJ,UK}
\author{Balvinder Paik}
\affiliation{Highams Park School, London, E4 9PJ,UK}
\author{Joanne Primrose}
\affiliation{Highams Park School, London, E4 9PJ,UK}
\author{Danny Quick}
\affiliation{Highams Park School, London, E4 9PJ,UK}
\author{Anais Radeka}
\affiliation{Highams Park School, London, E4 9PJ,UK}
\author{Anthony Rodney}
\affiliation{Highams Park School, London, E4 9PJ,UK}
\author{Eleanor Sandeman}
\affiliation{Highams Park School, London, E4 9PJ,UK}
\author{Fawad Sheikh}
\affiliation{Highams Park School, London, E4 9PJ,UK}
\author{Camron Stansfield}
\affiliation{Highams Park School, London, E4 9PJ,UK}
\author{Delayne Symister}
\affiliation{Highams Park School, London, E4 9PJ,UK}
\author{Joshua Taylor}
\affiliation{Highams Park School, London, E4 9PJ,UK}
\author{William Wilshere}
\affiliation{Highams Park School, London, E4 9PJ,UK}
\author{R. Wesson}
\affiliation{Dept. of Physics \& Astronomy, University College London, Gower St, London, WC1E 6BT,UK}
\author{I. De Looze}
\affiliation{Sterrenkundig Observatorium, Ghent University, Krijgslaan 281 - S9, 9000 Gent, Belgium}
\author{G. C. Clayton}
\affiliation{Department of Physics \& Astronomy, Louisiana State University, Baton Rouge, LA 70803, USA}
\author{K. Krafton}
\affiliation{Department of Physics \& Astronomy, Louisiana State University, Baton Rouge, LA 70803, USA}
\author{M. Matsuura}
\affiliation{School of Physics \& Astronomy, Cardiff University, Cardiff, Wales, UK}

\date{Accepted XXX. Received YYY; in original form ZZZ}

\label{firstpage}

\begin{abstract}
Core-collapse supernovae (CCSNe) are potentially capable of producing large quantities of dust, with strong evidence that ejecta dust masses can grow significantly over extended periods of time.
Red-blue asymmetries in the broad emission lines of CCSNe can be modelled using the Monte Carlo radiative transfer code {\sc damocles}, to determine ejecta dust masses. To facilitate easier use of {\sc damocles}, we present a Tkinter graphical user interface (GUI) running {\sc damocles}. The GUI was tested by high school students as part of the Original Research By Young Twinkle Students (ORBYTS) programme, who used it to measure the dust masses formed at two epochs in two Type~IIP CCSNe: SN 2012aw and iPTF14hls, demonstrating that a wide range of people can contribute significantly to scientific advancement. Bayesian methods were used to quantify uncertainties on our model parameters. From the presence of a red scattering wing in the day 1863 H$\alpha$ profile of SN~2012aw, we were able to constrain the dust composition to large (radius $>0.1~\mu$m) silicate grains, with a dust mass of $6.0^{+21.9}_{-3.6}\times10^{-4}~M_\odot$. 
From the day 1158 H$\alpha$ profile of SN~2012aw, we found a dust mass of $3.0^{+14}_{-2.5}\times10^{-4}$~M$_\odot$. For iPTF14hls, we found a day 1170 dust mass of 8.1 $^{+81}_{-7.6}\times10^{-5}$ M$_{\odot}$ for a dust composition consisting of 50\% amorphous carbon and 50\% astronomical silicate. At 1000 days post explosion, SN 2012aw and iPTF14hls have formed less dust than the peculiar Type II SN~1987A, suggesting that SN~1987A may have formed a larger dust mass than typical Type IIP's.

\end{abstract}

\maketitle



\section{Introduction}

The presence of large amounts of dust in high redshift young galaxies \citep[e.g.][]{Bertoldi2003, Watson2015a, Laporte2017b} challenged the longstanding assumption that most cosmic dust in the Universe was formed by AGB stars, which mostly form dust on long evolutionary timescales. It has long been known that dust can form in the ejecta of core-collapse supernovae (CCSNe) and \citet{Morgan2003a} and \citet{Dwek} have estimated that each CCSN would need to produce $\geq$0.1 M$_\odot$ of dust for CCSNe to be the main dust-producers in the Universe. Such yields have been predicted theoretically \citep[e.g.][]{Nozawa2003, Sarangi2015}. 
\\
\\
It is well-known that CCSNe can form large amounts of dust. Fits to the spectral energy distributions (SEDs) of warm dust emitting at mid-IR wavelengths as measured with the {\em Spitzer Space Telescope} found dust masses of only around $10^{-4}-10^{-2}$~M$_\odot$ \citep{Sugerman2006, Rho2009, Fabbri2011a}. However,  \citet{Matsuura} found a cold dust mass of $\sim$0.5~M$_\odot$ from measurements of the far-IR SED of SN~1987A taken with the {\em Herschel Space Observatory} 23 years after outburst. Subsequent analyses of {\em Herschel} far-IR data for Cas~A and other supernova remnants have detected large cold dust masses between 0.04 - 1.0~M$_\odot$ \citep{Gomez2012, DeLooze2017, Temim2017, Chawner2019, Priestley2019, DeLooze2019, Niculescu-Duvaz2021}.
\\
\\
In order to provide a method for measuring supernova dust masses that does not rely on infrared SED measurements, 
\citet{Bevan2016}, motivated by the work of \citet{Lucy1989} on SN~1987A, developed the Monte-Carlo radiative transfer code {\sc damocles}, which quantifies the amount of newly formed dust causing the absorption and scattering of CCSN ejecta line emission by modelling observed red-blue line asymmetries.
The effect is caused by light from the receding red-shifted side of a CCSN being attenuated by more internal dust than light from the approaching blue-shifted side. 

Many authors have noted the presence of red-blue asymmetries in the line profiles of CCSNe \citep{Smith2008, Mauerhan2012, Gall2014,Milisavljevic2012}. 
Dust mass measurements of SN~1987A between 714 and 3604 days by \citet{Bevan2016} using {\sc damocles} were in agreement with the work of \citet{Wesson2015} who modelled its IR SEDs at similar epochs, both finding that by day 3604 the dust mass in SN~1987A had reached $\sim$0.1 M$_{\odot}$, and that the dust mass increase with time could be fitted with a sigmoid function. {\sc damocles} has since been applied to other objects such as SN~1980K, SN~1993J, Cas~A
\citep{Bevan2017} and SN~2005ip \citep{Bevan2019}, with all these objects being found to have dust masses $>$0.1~M$_\odot$. 
\citet{Niculescu-Duvaz2022} used {\sc damocles} to model the multi-epoch optical spectra of 13 CCSNe aged between 5-60 years, compiling the most comprehensive set of very late-epoch CCSN ejecta dust measurements made to date. They confirmed that, on average, the dust mass growth could be fitted by a sigmoid curve, saturating on a timescale of $\sim$30 years at a value of 0.42$^{+0.09}_{-0.05}$~M$_\odot$.  
\\
\\
However, there are not yet enough CCSN dust mass estimates available to be able to discern correlations between dust mass, grain size and CCSN properties such as SN sub-type and progenitor mass. In this work, we aim to add to the sample of CCSNe that have derived dust masses with robustly quantified uncertainties across a range of epochs. We also try, where possible, to constrain the dust grain size and composition. These parameters can help constrain reverse shock dust destruction rates, which have been deduced to range between 0-100\% \citep{Nath, Silvia2010, Bocchio2014, Micelotta2016, Kirchschlager2019, Slavin2020, Priestley2021,Kirchschlager2020}, and which are strongly dependent on the properties adopted for the dust.
\\
\\
In this paper we present late epoch (days 752 to 1863) optical spectra of two Type~IIP CCSNe, SN~2012aw and iPTF14hls. Both show pronounced red-blue line asymmetries, with blue-shifted emission peaks attributable to internal dust attenuating more light from the red-shifted far side of the ejecta than from the blue-shifted near side.
From spectropolarimetry of SN~2012aw obtained between days 16-120 after outburst, a period when the optical line profiles still showed P~Cygni profiles, \citet{Dessart2021} deduced the presence of significant asymmetry in its ejecta. If ejecta asymmetries played a major role in shaping later epoch emission line profiles then we might expect to encounter equal numbers of CCSNe showing blue-shifted or red-shifted emission line peaks. In fact, the late-epoch spectral study by \citet{Niculescu-Duvaz2022}
found only blue-shifted emission line peaks amongst their sample. SN~1941C, observed in 2019 by \citet{Fesen2020}, is a rare example of a CCSN showing a net red-shifted late epoch emission line profile.
\\
\\
Section 2 describes the ORBYTS programme, through which part of this work was done, whilst Section 3 summarises the optical spectra of SN 2012aw and iPTF14hls modelled in this paper. There is no particular motivation behind the selection of these two CCSNe: iPTF14hls was found by students during an exercise searching through the wiserep archive to find objects with broad-line emission, and SN 2012aw was randomly chosen for students to model from the Gemini GMOS and Xshooter late-time CCSNe survey presented in Wesson et al. in prep 2022. it exhibited a fairly smooth and uncomplicated line model, which was ideal for testing the GUI Section 4 outlines the {\sc damocles} code and the methodology used to model the observed emission lines, whereas Section 5 describes a GUI wrapper written for {\sc damocles} to facilitate a more efficient and clear modelling process, which was deployed through the ORBYTS programme. Sections 6 and 7 present the results of our models of the H$\alpha$ line in SN 2012aw and iPTF14hls, and in Section 8 we present our conclusions.                                        

\section{Original Research By Young Twinkle Students (ORBYTS)}

ORBYTS is a National Educational Opportunities Network (NEON) award winning educational programme managed by Blue Skies Space Ltd. (BSSL) and University College London (UCL), in which secondary school pupils work on original research under the tutelage of PhD students and other young scientists \citep[][]{McKemmish2017a,Sousa2018}.
\\
\\
As part of the ORBYTS scheme, researchers pay fortnightly
visits to their partner school over the academic year and facilitate cutting-edge research. One of the core aims of the program is to address the under-representation of people from socially deprived backgrounds in Higher Education. Longitudinal schemes are the most effective way to increase pupil interest in taking STEM subjects at university \citep{simon2020}. The ORBYTS program achieves this goal by improving pupils' confidence in their abilities through regular interaction with their tutor, which changes the narrative on what it means to be a scientist.
\\
\\
ORBYTS spans a wide range of research areas, such as providing accurate molecular transition frequencies (\citep[][]{McKemmish2017b, McKemmish2018,Chubb2018b,Darby-Lewis2019}) and characterising the transit parameters of a range of exoplanets \citep[][]{Edwards2020,Edwards2021,Edwards2021b}. 
Other projects have included analysing CH$_3$OH and CH$_3$CHO emission in protostellar outflows \citet{Holdship2019}, applying of artificial intelligence to images \citep{francis2020}, X-ray spectral analysis of Active Galactic Nuclei \citep{Grafton-Waters2021} and planetary aurorae \citep{wibisono2020}.
\\
\\
In the current work, students partook in a wide range of research-related tasks, such as searching archival data for SNe with optical broad-line emission, as well as using the {\sc damocles} code to model line profiles in CCSNe, with the aim of quantifying parameters of the SN ejecta and the dust masses present.

\section{Observations}
The optical spectra of SN 2012aw we model in this work were taken with the 8-m Gemini-North telescope, and are part of a Gemini GMOS and VLT X-Shooter late-time spectroscopic survey of CCSNe presented by Wesson et al. (2022, in preparation). The Gemini GMOS-N spectra cover the range 4400-7500~\AA\ and were obtained on 18/05/2015 and 22/04/2017 in long-slit mode using the B600 grating, with a slit width of 0.75~arcsec. The spectra were taken at two or three central wavelength settings and co-added to prevent important spectral features from falling in detector gaps. The spectra have a resolution of 3.5~\AA\ at a wavelength of 6000~\AA . The 2D spectra were bias-corrected, flat-fielded and wavelength calibrated using the {\sc iraf} {\em gemini} package, and corrected for cosmic rays using the {\sc lacos} package of \citet{vanDokkum2001}.
The sky subtraction regions were determined by visual inspection and the spectra were extracted using 15 rows centered on the supernova's position. Plots of the Gemini spectra are shown in Figure~A1 of Appendix~A.
\\
\\
We retrieved two spectra of iPTF14hls from the WISeREP archive\footnote{\url{https://wiserep.weizmann.ac.il/}}. The first was taken on 08/11/2016 by \citet{sollerman2019} with the low-dispersion spectrograph FLOYDS on the LCO Haleakala (FTN) telescope. The spectrum has a wavelength range of 4000-9300~\AA\ and a resolution of 10~\AA . The second spectrum was taken on 14/01/2018 by \citet{sollerman2019} with the Low Resolution Instrument (LRIS) on the Keck~I telescope. The spectral range is 3070-10265~\AA , with a resolution of $\sim$5~\AA . More observational details of the spectra can be found in the referenced literature; plots of the two spectra are shown in Figure~A2 of Appendix~A.

\section{Methodology}

All models of the CCSNe presented in this work were created with the Monte Carlo radiative transfer code {\sc damocles}, which was initially presented and described in depth by \citet{Bevan2016}. {\sc damocles} is written in {\sc Fortran} 95 and parallelised with {\sc openmp} \citep{dagum1998}. It models the scattering and absorption of emission line photons by dust in an expanding ejecta. It has been benchmarked against analytic models of theoretical line profiles based on work by \citet{Gerasimovic1933a}, and also against numerical models of SN 1987A made by \citet{Lucy1989}. {\sc damocles} is a versatile code able to treat any arbitrary dust/gas geometry, as well as a range of velocity and density distributions and dust and gas clumping configurations. It can also handle a wide range of dust species and grain radii.
\\
\\
Our approach to modelling the line profiles is described for SN~1987A by \citet{Bevan2016}, and for a range of CCSNe by \citet{Niculescu-Duvaz2022}. The parameter space was first examined manually using a graphical user interface (GUI), which is described in Section \ref{sec:damoc-inter}, to find the best fitting model. All models had five free parameters: the outer expansion velocity V$_{max}$,the emitting shell radius ratio R$_{in}$/R$_{out}$, the density profile index $\beta$, where $\rho$ $\propto$ r$^{-\beta}$, the dust mass M$_d$ and grain radius $a$. We determined V$_{\rm max}$ from the point at which the observed flux vanishes on the blue-shifted part of the line profile, as the red wing of the profiles can be modified due  to dust scattering. R$_{\rm in}$/R$_{\rm out}$ was set so the model line profile matched the blue shifted peak in the observed profile and its matching red shifted inflection point. The density profile $\beta$ is identified from the gradient of the emission line profile wings. After these values are fixed, we then iterate over the grain radius and dust mass to fit the observed profile.
\\
\\
We assume that at the epochs considered here the line-emitting gas is optically thin, and that the emissivity distribution is proportional to the square of the local gas density. The gas is kept smoothly distributed for all models. As the modelled CCSNe are young, we assume that the supernova ejecta is in free expansion, such that V$_{max}$ = R$_{out}$/$t$, where $t$ is the age of the supernova. The dust and the gas are co-located in our simulations, such that the dust and the gas have the same V$_{\rm max}$ and $R_{\rm in}/R_{\rm out}$, and, for smoothly distributed dust models, the gas and the dust have the same $\beta$ values. The dust in our models consists of amorphous carbon (AmC), silicates, or a 50:50 mix of both. Occasionally, there can be a particularly pronounced red scattering wing in an observed line profile, which requires a dust species with an albedo $>$0.7 to produce an adequately fitting model. In this case, we can constrain the dust species in our model to be silicate dust with a grain radius between 0.1-1.0~$\mu$m, as only these grain configurations provide a high enough albedo. We use "astronomical silicate" as the silicate dust species, with the optical constants of \citet{draine1984ApJ...285...89D}, along with the BE AmC optical constants of \citet{zubko1996}.
\\
\\
We modelled the line profiles first using a smooth then a clumped dust distribution, where the radius of an individual clump is R$_{out}$/40. We prefer the dust masses derived for a clumped distribution over a smooth one. This is due to the fact that models of early observations of SN~1987A required dust to be present in optically thick clumps \citep{Lucy1989,Lucy1991,Bouchet1996,Kozma1998b}. Similarly to the work of \citet{Niculescu-Duvaz2022}, we fixed the clump distribution power law index to be 3 and the clump volume filling factor to be 0.1.
We note that clumped dust models display less of a scattering wing and attenuate the red wing of the line profile less than smooth dust models, given no change in the other model parameters between the two cases. When the dust is located in clumps, impacting radiation is subject to less scattering as well as to less absorption. So, for a clumped dust model to match the line profile generated by a smooth dust model, a small modification to the grain radius to produce a larger albedo, as well as a dust mass larger 
by a factor of $\sim$2 than for the smooth case, are required.
\\
\begin{figure*}
\centering

\includegraphics[width=0.9\linewidth]{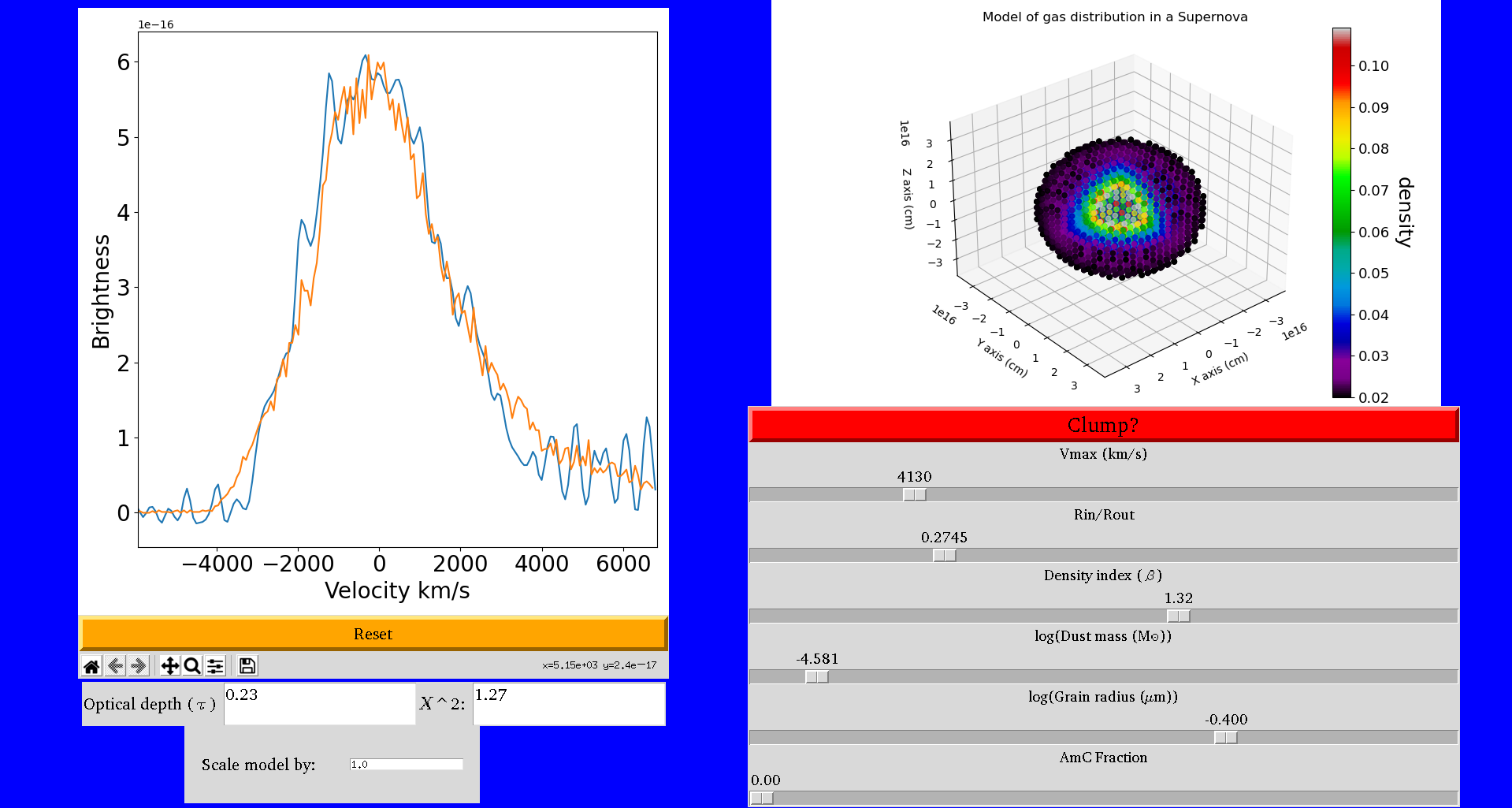}

\caption{The Tkinter GUI used to explore the parameter space of {\sc damocles} models in this work. }
\label{fig:damoc-gui}
\end{figure*}

We conducted a Bayesian analysis on every emission line profile modelled in this work, so as to quantify errors on the model parameters. We use the Bayesian approach to modelling line profiles with {\sc damocles} first described by \citet{Bevan2018}. We use an affine invariant Markov Chain Monte Carlo (MCMC) ensemble sampler, {\sc emcee} \citep{Goodman2010, Foreman-Mackey2013} with {\sc damocles} to sample the posterior probability distribution of the input parameters. This is defined by Bayes' Theorem, presented in equation (\ref{eq:bayeseq}): 
\begin{equation} \label{eq:bayeseq}
P(\theta|D) \propto P(\theta)P(D|\theta)
\end{equation}
In this equation, D is the data, $\theta$ is the set of parameters of the model, P($\theta$) is our prior understanding of the probability of the parameters, and P(D|$\theta$) is the likelihood, which is the probability of obtaining the data for a given set of parameters.
The likelihood function we used is proportional to $\exp{-\chi^2_{red}}$, where $\chi^2_{red}$=$\chi^2$/$\upsilon$, and $\upsilon$ is the number of degrees of freedom and $\chi^2$ is defined by equation (\ref{eq:likelihoodeq}):
\begin{equation} \label{eq:likelihoodeq}
\chi^2 = \sum^{n}_{i=1}
\frac{(f_{mod,i} - f_{obs,i})^2}{\sigma_i^2} 
\end{equation}
where f$_{mod,i}$ is the modelled flux in bin $i$, f$_{obs,i}$ is the observed flux in frequency bin $i$, and $\sigma_i$ is the combined Monte Carlo and observational uncertainty in bin $i$. The modelled line profile from which f$_{mod,i}$ is taken is normalised to the peak flux of the observed line profile. Priors for each model are given in uniform space, with the exception of the dust mass and grain radius, which are given in log-uniform space as their possible values can span several magnitudes.
\\
\\
Final posterior probability distributions for our input parameters are presented as a series of 2D contour plots, where each pair of parameters are marginalised over the other parameters. We also present a 1-D
marginalised posterior probability distribution for each parameter. The "best fitting" parameter value found from the Bayesian analysis is given as the median of the marginalised 1-D probability distribution. The mean was not used as many probability distributions deviated from a Gaussian distribution. The lower and upper limits were taken to be the 16th and 84th quartiles for the 1-D posterior probability distributions. The Bayesian model fits presented in this work all use a 100~per~cent clumped silicate dust composition. In the case of iptf14hls, to find uncertainties on the dust mass for a 50:50 AmC to silicate dust ratio, we find the percentage errors on the manually-determined dust mass from the limits determined by the Bayesian analysis for a 100~per~cent silicate composition and scale them to that value.
\\
\\
As we had to run many models, we restricted our parameter space to 5 dimensions, representing the parameters described above. Our parameter space was explored by 250 walkers. For each parameter, the number of iteration steps for the autocorrelation function to initially decay down towards zero is roughly one autocorrelation time. We checked that each model was run for 5 or more autocorrelation times to ensure convergence. The acceptance fraction is the fraction of times the emcee algorithm has accepted a new proposed set of values. An acceptable value of the acceptance fraction is between $0.2-0.5$ \citep{Gelman1996}. For each simulation our acceptance fraction  averaged at around 0.3. 
\\

\section{The interactive {\sc DAMOCLES} environment}\label{sec:damoc-inter}

Here we describe the GUI we created, which was tested through the ORBYTS programme where students used it to find the best fitting {\sc damocles} models of the H$\alpha$ line profiles of SN~2012aw and iPTF14hls. This interactive environment provides a user-friendly way to manually explore parameter space and investigate how each parameter affects the input model. It has been made public on GitHub \footnote{\url{https://github.com/damocles-code/damocles}}. 
It is written in Python using the Tkinter package, and uses the Fortran to Python interface generator F2PY3 to run the damocles code from the GUI. Upon running the GUI, the user is first prompted to enter relevant information, for example the wavelength of the emission line of interest and the filename and epoch of the observed data. Upon completion of the necessary fields, the main window where {\sc damocles} models are created is spawned. The layout of this window can be seen in Figure \ref{fig:damoc-gui}. The user sets the parameters of the gas and dust CCSN model via sliders. There are sliders for V$_{max}$, R$_{in}$/R$_{out}$, the density profile index $\beta$, the dust mass (M$_d$), grain radius ($a$) and the fraction of AmC to silicate dust. When a slider is moved to a desired value, the current values of each slider is set in the input files of {\sc damocles}. The user can push the red button above the sliders if they want to switch from a smooth dust distribution to a clumped one. The default filling factor for the clumped dust model is 0.1, where the power law index for the clump number density distribution is 3.0. The model 3D gas shell for the CCSN is then generated and plotted in the upper right pane. This is useful for visual purposes, as it informs the user of how changing a parameter affects the distribution of gas in the model. {\sc damocles} is called to run within the GUI using the parameters specified by the user. The output line profile generated by {\sc damocles} is overplotted onto the observed line profile in the left-hand pane. The $\chi^2$ value between the model and observed line profile is provided below this pane, as is the radial optical depth, $\tau$, of the dust shell between R$_{in}$ and R$_{out}$.

\section{SN 2012aw}
SN~2012aw in the galaxy M~95 was discovered on March 16th 2012 by \citet{Fagotti2012}. Observations of M~95 a day earlier \citep{poznanski2012}, with a limiting magnitude of R $>$ 20.7, did not detect the supernova, so the explosion date can be well constrained to March 16th 2012.
\citet{Bose2013} classified it as Type~IIP from fitting the light curve of the archetypal Type~IIP SN~1999em to photometric points taken 4 - 270 days after the explosion date. The relative proximity of M~95, as well as the relatively low dust extinction towards it, meant that SN 2012aw was the first supernova to have a plateau detected in its ultraviolet lightcurve \citep{Bayless2013}.
\\
\\
\begin{figure*}
\centering

\includegraphics[width=0.9\linewidth]{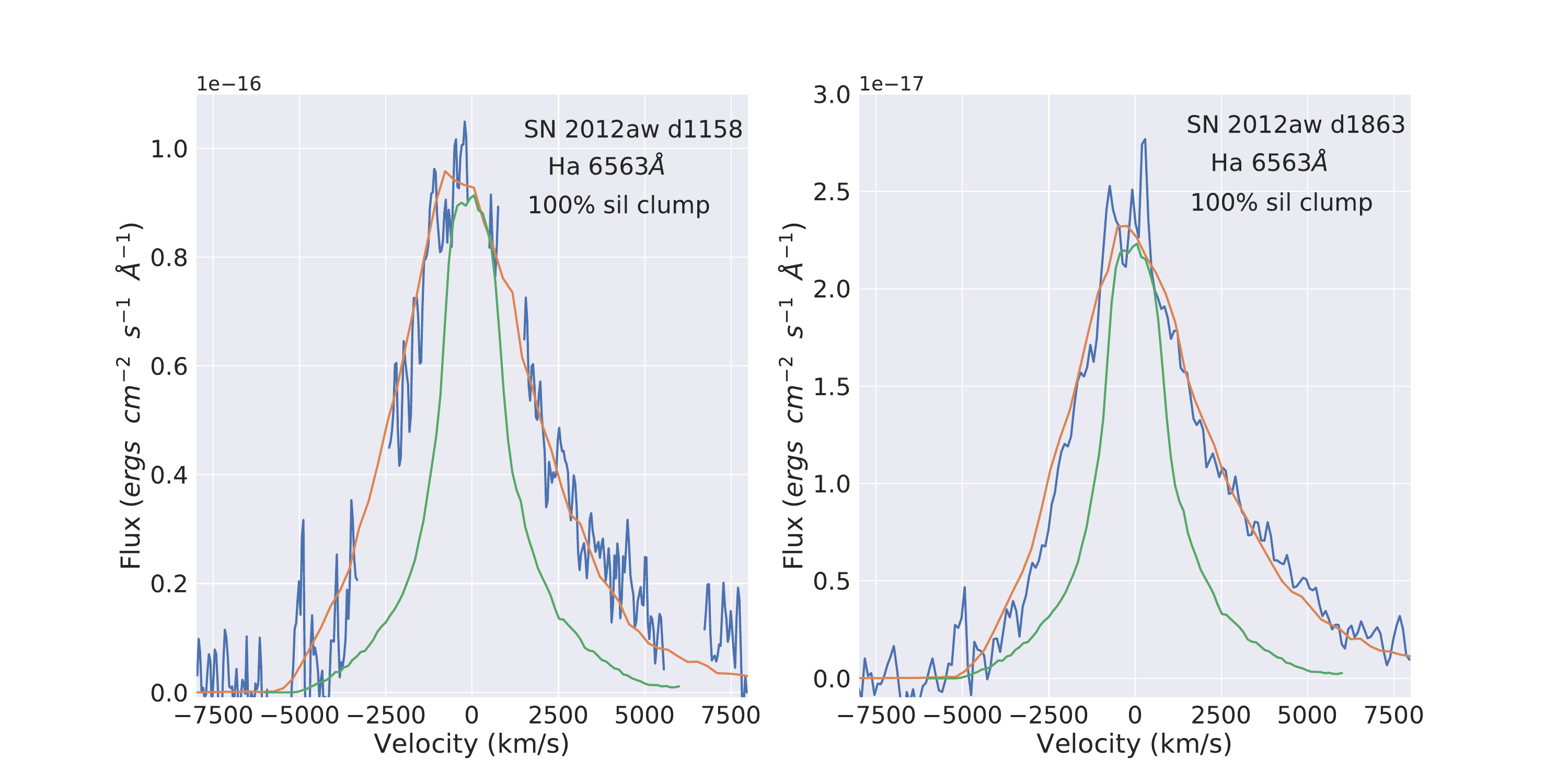}

\caption{Dust-affected {\sc damocles} models for the H$\alpha$ profile (orange line) of SN~2012aw in Gemini GMOS spectra (blue) taken 1158 days (left) and 1863 days (right) post-explosion. The green line is the dust-free model, normalised to the peak level of the observed profile. Clumped dust models corresponding to 100\% astronomical silicate are shown.}
\label{fig:2012aw-fits}
\end{figure*}

\begin{figure*}
\centering

\includegraphics[width=\linewidth]{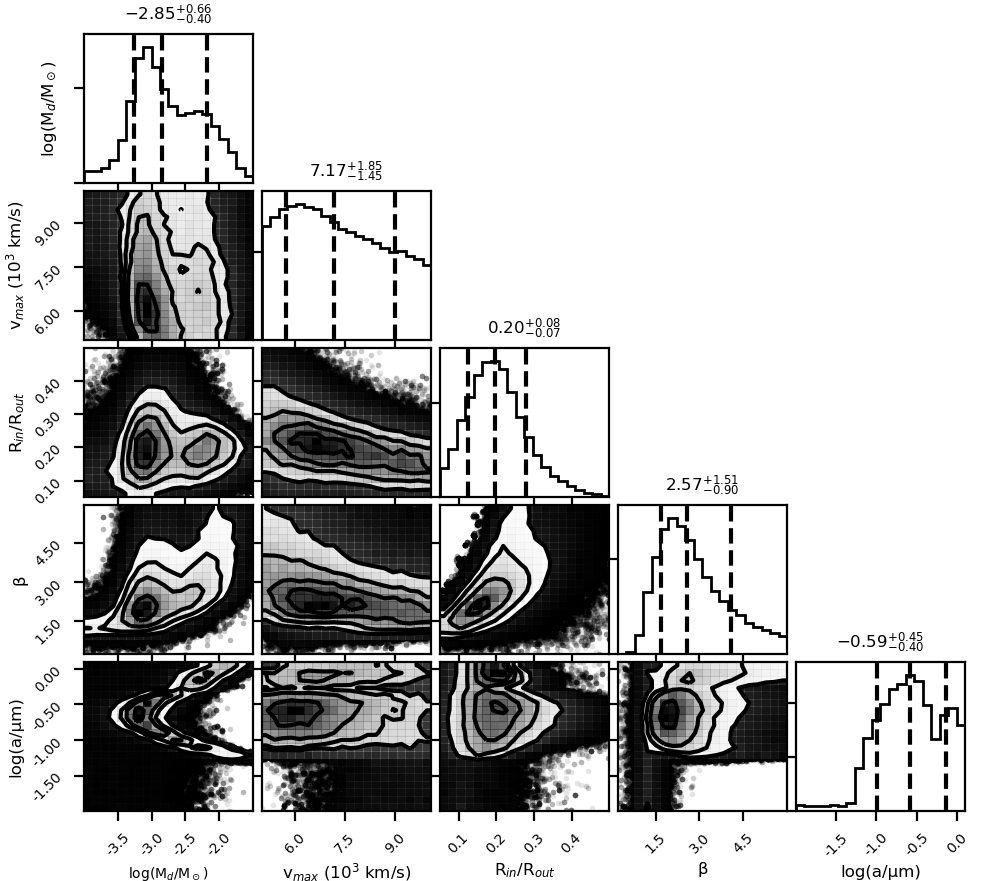}

\caption{The full posterior probability distribution for the 5D model of SN~2012aw's H$\alpha$ profile at day 1863, using a 100\% silicate dust composition. The contours of the 2D distributions represent 0.5$\sigma$, 1.0$\sigma$, 1.5$\sigma$ and 2.0$\sigma$ and the dashed, black vertical lines represent the 16th, 50th, and 84th quartiles for the 1-D marginalized probability distributions.}
\label{fig:bayesian-2012aw}
\end{figure*}

\begin{table*}
\centering
\caption{
Parameters used in the {\sc damocles} models, found from a manual examination of parameter space, of the broad emission lines of SN~2012aw and iPTF14hls for spherically symmetric smooth and clumped dust models. The "\% Sil" column stands for percentage of the dust species that is astronomical silicate, where the remainder of the dust is AmC dust. The optical depth $\tau$ is calculated from R$_{in}$ to R$_{out}$ for a central line wavelength for H$\alpha$ of 6563~\AA. The columns labelled $+$ and $-$ represent the upper and lower uncertainties on the dust mass, such that the maximum possible dust mass is M$_dust$ added to the corresponding value listed the $+$ column, and the minimum possible dust mass is the corresponding value in the $-$ column subtracted from M$_dust$. These limits are derived from the 16th and 84th quartiles on the median value of the 1D posterior PDF for the dust mass found from a Bayesian simulation, and are scaled as a percentage error to the reported values of M$_{dust}$ in this table.}
\begin{tabular}{ccccccccccccccc}
\hline
SN & Line & Epoch & Clumped & \% Sil & a & V$_{max}$ & V$_{min}$ & $\beta_{gas}$ & R$_{out}$ & R$_{in}$ & M$_{dust}$ & + & - & $\tau$  \\
 & & days & & & $\mu$m & km~s$^{-1}$ & km~s$^{-1}$ & & 10$^{13}$~cm & 10$^{13}$~cm & $10^{-4}$~M$_\odot$ & $10^{-4}$~M$_\odot$ & $10^{-4}$~M$_\odot$ & \\
\hline
2012aw & H$\alpha$         & 1158 & no  & 100 & 0.30  & 5500  & 1100 & 1.30 & 55.03  & 11.01 & 2.0 & 8.9 & 1.6 & 0.70  \\
2012aw & H$\alpha$         & 1158 & yes & 100 & 0.30  & 5500  & 1100 & 1.30 & 55.03  & 11.01 & 2.0 & 8.9 & 1.6 & 0.70  \\
2012aw & H$\alpha$         & 1158 & no  & 75  & 0.25 & 5500  & 1100 & 1.30 & 55.03  & 11.01 & 1.5 & 6.6 & 1.2 & 0.64  \\
2012aw & H$\alpha$         & 1158 & yes & 75  & 0.25 & 5500  & 1100 & 1.30 & 55.03  & 11.01 & 1.0 & 4.4 & 0.8 & 0.43  \\
2012aw & H$\alpha$         & 1863 & no  & 100 & 0.30  & 5100  & 663  & 1.53 & 82.09  & 10.67 & 7.0 & 25.5 & 4.2 & 1.19  \\
2012aw & H$\alpha$         & 1863 & yes & 100 & 0.30  & 5100  & 663  & 1.53 & 82.09  & 10.67 & 6.0 & 21.8 & 3.6 & 1.02  \\
2012aw & H$\alpha$         & 1863 & no  & 75  & 0.25 & 5100  & 663  & 1.60 & 82.09  & 10.67 & 5.0 & 18.2 & 3.0 & 1.03   \\
2012aw & H$\alpha$         & 1863 & yes & 75  & 0.25 & 5100  & 663  & 1.60 & 82.09  & 10.67 & 5.0 & 18.2 & 3.0 & 1.03  \\
iPTF14hls & H$\alpha$ & 752 & yes & 0   & 0.18  & 4000 & 1107 & 1.30 & 27.2 & 7.5  & 0.10 & 1.20 & 0.096 & 0.34  \\
iPTF14hls & H$\alpha$ & 752 & no  & 0   & 0.15  & 4000 & 1107 & 1.30 & 27.2 & 7.5  & 0.03 & 0.37 & 0.029 & 0.12 \\
iPTF14hls & H$\alpha$ & 752 & yes & 100 & 0.044 & 4100 & 1107 & 1.13 & 27.2 & 7.5  & 0.28 & 3.5 & 0.27 & 0.26  \\
iPTF14hls & H$\alpha$ & 752 & no  & 100 & 0.057 & 4100 & 1107 & 1.30 & 27.2 & 7.5  & 0.21 & 2.60 & 0.20 & 0.22 \\
iPTF14hls & H$\alpha$ & 752 & yes  & 50 & 0.10 & 4100 & 1107 & 1.30 & 27.2 & 7.5  & 0.13 & 0.12 & 1.60 & 0.23 \\
iPTF14hls & H$\alpha$ & 752 & no  & 50 & 0.09 & 4100 & 1107 & 1.30 & 27.2 & 7.5  & 0.18 & 0.17 & 2.2 & 0.32 \\
iPTF14hls & H$\alpha$          & 1170 & yes & 0   & 0.18 & 6025 & 1085 & 2.30 & 60.90 & 10.96 & 0.47 & 4.7 & 0.44 & 0.34  \\
iPTF14hls & H$\alpha$          & 1170 & no  & 0   & 0.15 & 6025 & 1085 & 2.30 & 60.90 & 10.96 & 0.46 & 4.6 & 0.43 & 0.40  \\
iPTF14hls & H$\alpha$          & 1170 & yes & 100 & 0.044 & 6025 & 1085 & 2.30 & 60.90 & 10.96 & 18.0 & 179.4 & 16.9 & 0.28  \\
iPTF14hls & H$\alpha$         & 1170 & no  & 100 & 0.038 & 6025 & 1085 & 2.30 & 60.90 & 10.96 & 18.6 & 174.3 & 17.4 & 0.34  \\
iPTF14hls & H$\alpha$         & 1170 & yes  & 50 & 0.10 & 6025 & 1085 & 2.30 & 60.90 & 10.96 & 0.81   & 8.1 & 0.76 & 0.33  \\
iPTF14hls & H$\alpha$         & 1170 & no  & 50 & 0.10 & 6025 & 1085 & 2.30 & 60.90 & 10.96 & 0.61   & 6.1 & 0.57 & 0.26  \\
\end{tabular}
\label{table:model-params}
\end{table*}

\begin{figure*}
\centering

\includegraphics[width=0.9\linewidth]{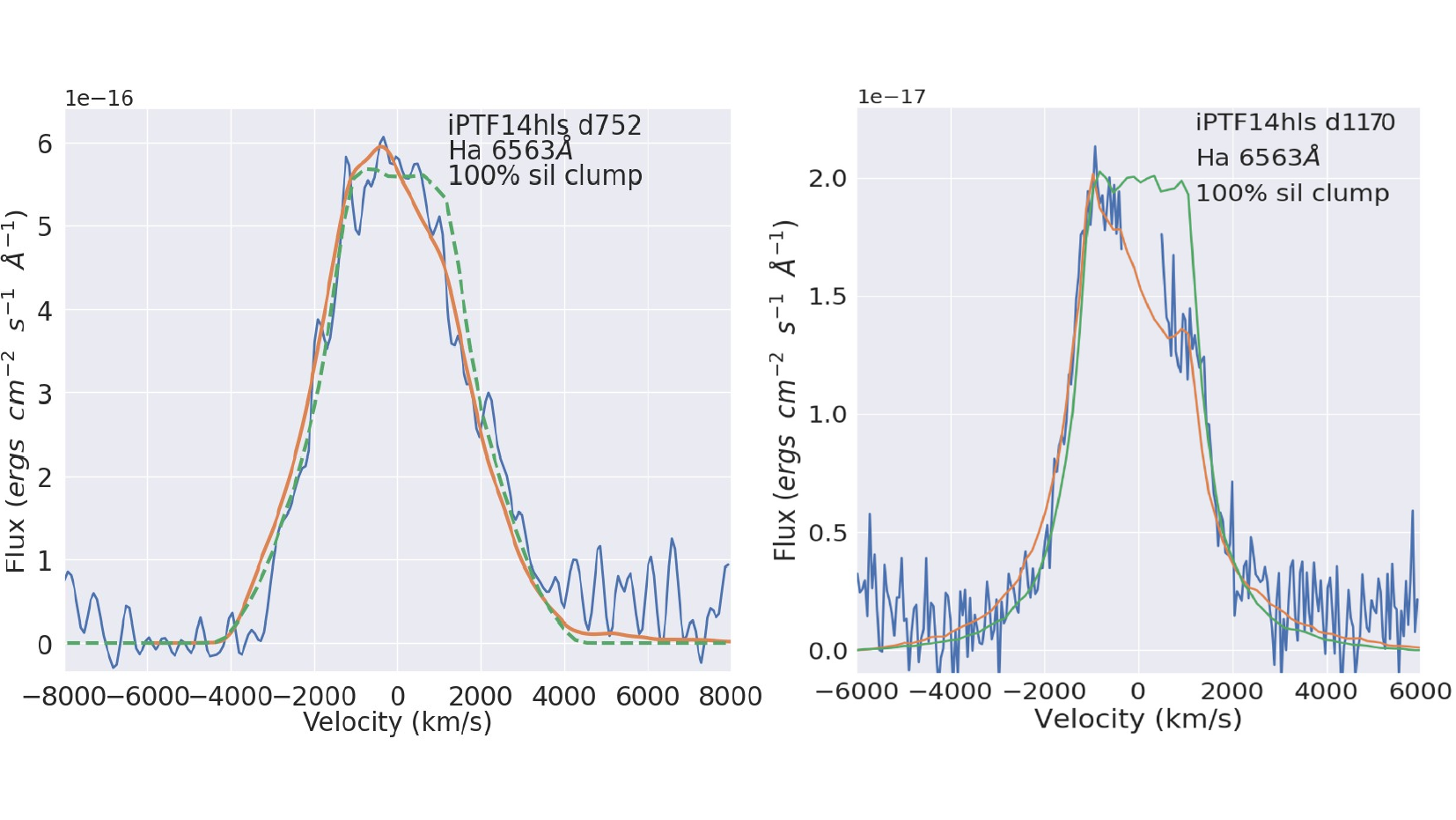}

\caption{Dust-affected {\sc damocles} models (orange lines) for the observed (blue) H$\alpha$ line profiles of iPTF14hls at 752 days (left) and 1170 days (right) post-explosion. The green lines are the dust-free models, normalised to the peak levels of the observed profiles. Clumped dust models corresponding to 100~per cent astronomical silicate are shown.}
\label{fig:iptf14-fits}
\end{figure*}

\begin{figure*}
\centering
\includegraphics[width=\linewidth]{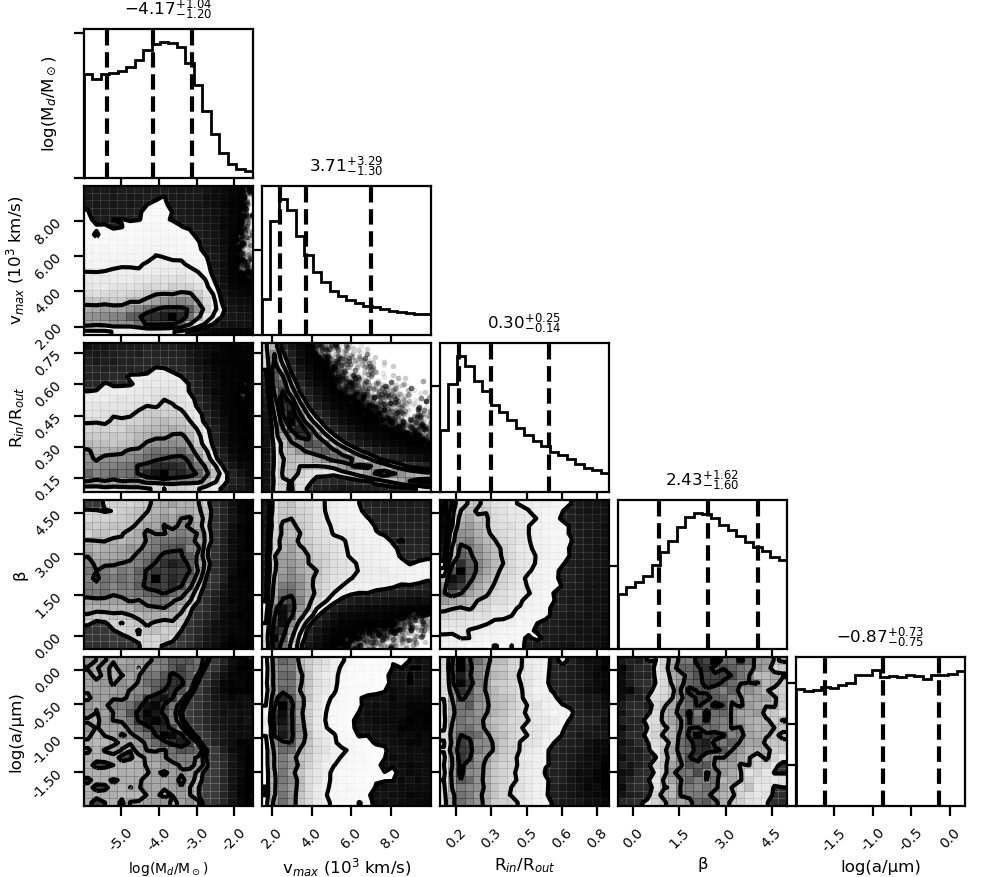}

\caption{The full posterior probability distribution for the 5-D model of the iPTF14hls H$\alpha$ profile at day 1170, using a 100\% clumped silicate dust composition. The contours of the 2D distributions represent 0.5$\sigma$, 1.0$\sigma$, 1.5$\sigma$ and 2.0$\sigma$ and the dashed, black vertical lines represent the 16th, 50th, and 84th quantiles for the 1-D marginalized probability distributions.}
\label{fig:bayesian-iptf}
\end{figure*}

The progenitor mass of SN~2012aw has been estimated by several authors. \citet{VanDyk2012} detected the near-IR signal of a progenitor in archival data taken 6-12 years before explosion, as well as an optical signal 17-18 years before, from archival {\em HST} data. They estimated that the progenitor was likely a red supergiant, with a progenitor mass of 17-18~M$_\odot$. 
from fitting the observed IR SED with synthetic photometry from a MARCS model stellar atmosphere, then comparing the derived T$_{eff}$ and luminosity to theoretical massive-star evolutionary tracks. 
They also estimated a high visual extinction in front of the progenitor star (A$_{\rm V}$ = 3.1~mag), much higher than the post-explosion estimate (A$_{\rm V}$ = 0.24 mag), which they interpreted as implying the destruction of a large amount of circumstellar dust by the explosion. \citet{Jerkstrand2014} obtained optical spectra 250-451 days after explosion and near-IR spectra 306 days after explosion.
From spectral synthesis models to their data they constrained the progenitor mass of SN~2012aw to be 14-18~M$_\odot$. \citet{Fraser2016} independently estimated the progenitor mass to be $12.5\pm1.5$~M$_\odot$.
\\
\\
ORBYTS pupils used the GUI applet described in Section \ref{sec:damoc-inter} to find a best-fitting model the H$\alpha$ line profile in our GMOS spectra of SN~2012aw taken at 1158 and 1863 days past explosion, by reducing the $\chi^2$ value. The spectra were corrected for an M~95 recessional velocity of 779~km~s$^{-1}$ \citep{Springob2005}. Models were convolved to the spectral resolution of the GMOS instrument using a gaussian kernel. 
Best-fitting models are shown in Figure~\ref{fig:2012aw-fits} and the model parameters are listed in Table~\ref{table:model-params}.
Models made using amorphous carbon as the dust grain species failed to replicate the pronounced red scattering wing, and at both epochs and we were unable to achieve a good fitting model to the H$\alpha$ profile using a silicate to carbon ratio of less than 0.75. From a manual fitting process we were able to constrain the grain radius at both epochs to be $\sim0.3~\mu$m for a 100\% silicate dust composition.
\\
\\
We quantified the uncertainties on the input parameters at both epochs using a Bayesian analysis with a 100\% silicate dust composition, the corner plot for which can be seen in Figure \ref{fig:bayesian-2012aw}. Grain radii smaller than 0.1~$\mu$m were strongly ruled out, while the median grain radius of 0.25$^{+0.47}_{-0.15}$~$\mu$m that was found was very close to the value of 0.3~$\mu$m found from manually fitting the line profile, while the dust masses we determined manually of $6.0^{+21.9}_{-3.6}\times10^{-4}~M_\odot$ also fell within the limits determined by the Bayesian analysis.

From a Bayesian analysis we found a lower median dust mass for SN~2012aw at day 1158, $3.0\times10^{-4}$~M$_\odot$, very close to our initial estimate of $2.0^{+8.9}_{-1.6}\times10^{-4}$~M$_\odot$, but the grain size was not as well constrained at this earlier epoch. These dust masses have been plotted in the dust mass growth with time plot shown in Figures 22 and 23 of \citet{Niculescu-Duvaz2022}.

\section{iPTF14hls}

iPTF14hls was discovered on September 22nd 2014 by the intermediate Palomar
Transient Factory (iPTF) wide-field camera survey \citep[][]{arcavi2017}. We adopt their discovery date as the explosion date, as well as the redshift of z = 0.0344 they determined from narrow host-galaxy features. 
iPTF14hls was classified as a supernova of Type IIP by \citet{Li2015} from the
broad P-Cygni Balmer series lines in an optical spectrum taken on January 8th 2015. iPTF14hls is an unusual supernova: peaking at $r=-19.1$ \citep[][]{sollerman2019}, its light curve as observed by \citet{arcavi2017} showed five distinct peaks over the first 600 days after explosion. The spectra resembled those of other hydrogen-rich supernovae (SNe) such as SN~1999em, but notably evolved at a much slower pace. \citet{sollerman2019} continued to monitor the light curve until day 1200, and noted a steep decline in the lightcurve around day 1000. They also obtained several optical spectra between day 713 and 1170. \citet{Yuan2018} detected a variable $\gamma$-ray source with the Fermi Large Area Telescope at a time and position consistent with iPTF14hls, and follow-up work by \citet{Prokhorov2021} strengthened this association. 
\\
\\
A number of authors have attempted to interpret the mechanism inducing the properties exhibited by iPTF14hls. \citet{arcavi2017} and \citet{Chugai2018} suggested iPTF14hls to have had a massive progenitor which prior to explosion underwent extreme mass loss, possibly caused by the pulsational pair-instability mechanism. \citet{Dessart2018} reproduced most of the observed properties of iPTF14hls with their model of a typical Type-II supernova created by a H-rich blue supergiant explosion and powered by a magnetar. \cite{Andrews2018} interpreted the day 1153 optical spectrum as showing clear signs of CSM-ejecta interaction, and proposed a set-up for iPTF14hls where an asymmetric disc or torus of circumstellar material (CSM) was overrun and hidden below the photosphere of the expanding supernova ejecta, whereby variations in the density structure of the CSM could explain the multiple peaks in the light curve. \citet{Moriya2020} and \citet{Uno2020} proposed that iPTF14hls was not a supernova, but that it could have been a $\sim$100~M$_{\odot}$ star experiencing variable mass-loss episodes similar to $\eta$~Carinae, or a binary system of two $\sim$100 M$_{\odot}$ stars experiencing a dynamical common-envelope evolution. \citet{soker2018} posited that iPTF14hls consisted of a binary system where a neutron star spiraling inside a massive stellar envelope accreted mass and launched jets which ejected the circumstellar shell and eventually resulted in a final SN explosion.
\\
\\
The H$\alpha$ line profile in iPTF14hls was modelled by ORBYTS pupils at two epochs (752 and 1170 days after explosion) using {\sc damocles}. We do not model the [Ca~{\sc ii}] 7330~\AA\ doublet, despite a strong detection at day 1170, as there could be potential contamination from the [O~{\sc ii}] 7319,7330~\AA\ doublet. Whilst the [O~{\sc i}] 6300,6363~\AA\ line is also visible at this epoch, we considered it to have too low a signal to noise ratio to model. 
All models were convolved using a gaussian kernel to the spectral resolution of the instrument used to collect the observed data. Contaminating narrow line emission was removed from the  H$\alpha$ line profile.
The day 752 and day 1170 H$\alpha$ line profile fits, determined from a manual exploration of parameter space, are shown in Figure \ref{fig:iptf14-fits}. The model parameters for both epochs can be found in Table \ref{table:model-params}. 
\\

\begin{figure*}
\centering

\includegraphics[width=0.9\linewidth]{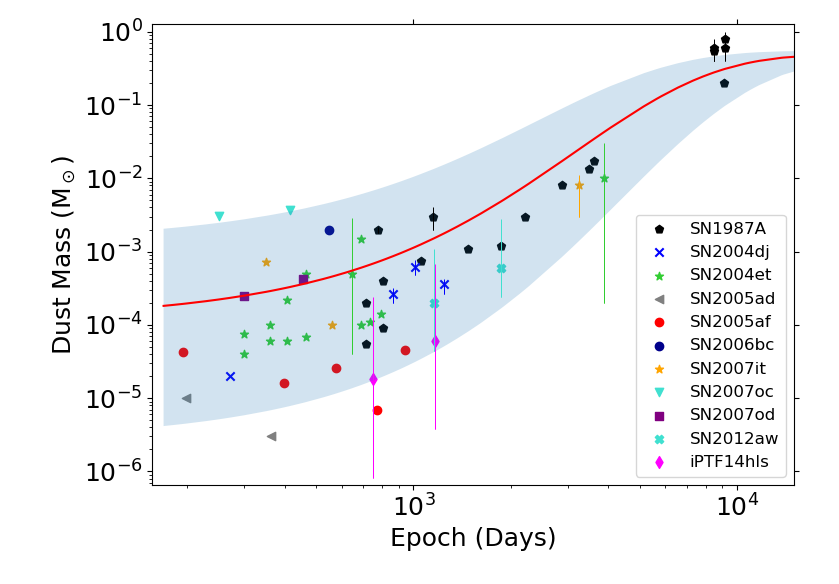}

\caption{A plot of dust mass versus time since explosion for Type~IIP CCSNe with dust mass measurements. Datapoints for SN~1987A are taken from \citet{Matsuura,Indebetouw2014,Matsuura2015,Wesson2015,Bevan2016}, for SN~2004et from \citet{Kotak2009,Fabbri2011a,Niculescu-Duvaz2022} and for SN~2007it also from \citet{Niculescu-Duvaz2022}. The datapoints for SN~2007oc are from \citet{szalai2013}, from \citet{andrews2010} for SN~2007od, from \citet{szalai2011} for SN~2004dj and from \citet{szalai2013} for SN~2005ad and SN~2005af. For SN~2006bc the data are taken from \citet{gallagher2012}. The red sigmoid curve is taken from Fig 23. of \citet{Niculescu-Duvaz2022}, where a dust growth trend is fit to a large sample of CCSNe dust mass measurements. The grey band encloses the error region on the best-fitting sigmoid curve parameters, where the errors are derived from a Monte Carlo bootstrap simulation.}
\label{fig:iip-compar}
\end{figure*}

Bayesian simulations were run for both epochs to constrain the uncertainty limits on the dust mass. The resulting corner plot for the Bayesian simulation of the H$\alpha$ profile at day 1170 can be found in Figure \ref{fig:bayesian-iptf}. At both epochs, the median dust mass found by a Bayesian exploration of parameter space and the best-fitting dust mass which minimized the $\chi^2$ found by a manual-fitting process are within the uncertainty limits derived from the 1D dust mass posterior probability distribution, for a 100\% clumped silicate dust distribution. As the dust species could not be determined, we report a final dust mass at day 1170 of 8.1 $^{+81}_{-7.6}\times10^{-5}$ M$_{\odot}$ for a 50:50 AmC to amorphous silicate dust ratio, both fixed with a dust grain radius of 0.1$\mu$m which is required to achieve a better fit to the line profile. At both epochs, the lack of a red scattering wing and the relatively low signal to noise of the H$\alpha$ line profiles meant that we were unable to constrain the dust grain radius. Therefore, the dust mass derived for this grain radius can be seen as a lower limit, as smaller silicate grains than 0.1$\mu$m have a poor extinction efficiency, as do larger carbon grains, such that smaller or larger grains require more dust to produce the same optical depth.
\\
\\
The ejecta dust mass found for SN~1987A at day 1153 from modelling the optical-infrared SED with a clumped AmC dust distribution was ($3.0\pm1.0)\times10^{-3}$~M$_\odot$ \citep[][]{Wesson2015}, while \citet{Bevan2016} 
modelled the day 1054 red-blue asymmetries of the [OI] 6300,6363A doublet using a clumped 100\% AmC composition to obtain a dust mass of $3.0^{+2.0}_{-2.25}\times10^{-3}$~M$_{\odot}$, nearly two orders of magnitude larger than our 100\% AmC dust mass of 4.7$\times10^{-5}$ M$_{\odot}$ for iPTF14hls at day 1170, and outside of the upper limit of 9.4$\times10^{-5}$ M$_{\odot}$ determined from our iPTF14hls measurement.
\\
\\
For our day 752 models for a 50:50 sil:AmC dust mixture and a grain radius of 0.1 $\mu$m, we find a very small amount of dust: 1.0$^{+14}_{-1.24}\times10^{-5}$ M$_{\odot}$. Although consistent with an increase in dust mass between days 752 and 1170, and despite the appearance of an increasing red-blue asymmetry in the H$\alpha$ line profiles between those dates, the uncertainties on the dust masses found using Bayesian methods are large enough that we are unable to definitively conclude that the dust mass has grown between these two epochs. 

The $\chi^2$ for the best-fitting 100\% silicate dust distribution provides a marginally better fit than that of a 100\% amorphous carbon distribution, but the low signal to noise of the spectrum means it is difficult to make out the continuum level, such that the perceived scattering wing better fit by a 0.1$\mu$m silicate grain could in fact be due to the unclear continuum.

\section{Discussion and Conclusions}
We have used the Monte Carlo radiative transfer code {\sc damocles} to model the
late-epoch H$\alpha$ line profiles 
of two Type~IIP supernovae, SN~2012aw and iPTF14hls, in order to determine their ejecta dust masses and grain parameters, as well as other ejecta parameters, for epochs corresponding to days 1158 and 1863 for SN~2012aw and to days 752 and 1170 for iPTF14hls.
We present a GUI wrapper for the {\sc damocles} code, written using the Python Tkinter module which facilitates a visual understanding of the code and a quick determination of the best fitting model to a broad emission line profile for a range of dust compositions. It can be used by both researchers and for citizen science purposes.
We have used this GUI to model the H$\alpha$ red-blue line asymmetries of the two supernovae. We have robustly quantified the errors on the derived model parameters using Bayesian inference. 
\\
\\
For SN~2012aw we constrained the day~1863 dust composition to be $>$75~per~cent silicate, and constrained the grain radius to 0.26$^{+0.46}_{-0.16}$~$\mu$m. Using this dust composition the dust mass found for a 100~per~cent silicate composition at day 1158 post-explosion was $2.0^{+8.9}_{-1.6}\times10^{-4}$~M$_\odot$,
while at day 1863 it was measured to be $6.0^{+21.9}_{-3.6}\times10^{-4}~M_\odot$. The large grain radius that we determine for SN~2012aw lends extra weight to the findings of \citet{Niculescu-Duvaz2022} that dust grains in most CCSNe have a grain radius of $>$0.1~$\mu$m, which is also in agreement with other studies constraining dust grain sizes (e.g. \cite{Gall2014}, \cite{owen2015},\cite{Wesson2015}, \cite{Bevan2017} and \cite{priestley2020}). Larger dust grains are more likely to survive the passage of the reverse shock than smaller grains, with \citet{Slavin2020} and \citet{Kirchschlager2020, Kirchschlager2022} finding a reverse shock dust destruction rate for silicate grain radii $>0.1~\mu$m of $\sim$20-50~per~cent.
\\
\\
For iPTF14hls at day 752 we found a dust mass of 1.0$^{+14}_{-1.24}\times10^{-5}$ M$_{\odot}$, while at day 1170 we found a dust mass of 8.1$^{+81}_{-7.6}\times10^{-5}$ M$_{\odot}$ for a dust composition of 50~per~cent amorphous carbon and 50~per~cent astronomical silicate. The large uncertainty limits on the dust mass measurements for iPTF14hls, and the fact that we could not constrain its dust grain composition, means we are unable to be certain that it had formed less dust than SN~2012aw at an epoch of $\sim$1200 days.
\\
\\
As SN~2012aw and iPTF14hls are fairly bright and young CCSNe, it should be straight-forward to conduct follow-up optical observations to monitor the evolution of their dust mass, as they are predicted to have only formed less than 1~per~cent of their total dust mass by 1000 days post-explosion (see Figure 23 of \citet{Niculescu-Duvaz2022}). An ongoing comparison between the evolution of these CCSNe and SN 1987A would be particularly interesting, as they are all classified as Type IIP CCSNe, but at day $\sim$1200 both SN~2012aw and iPTF14hls have been found here to have formed less dust than SN~1987A, as can be seen in Figure \ref{fig:iip-compar}. At around day 1000 post-explosion, the dust mass measured in SN 2004dj is also less than that found in SN 1987A. This hints that at 1000 days the peculiar Type~II SN~1987A could be an anomalously high dust-former, but more Type IIP SN measurements would need to be made to confirm this for later stages.
\\


\section*{Acknowledgements}

MND, MJB and RW acknowledge support from European Research Council (ERC) Advanced Grant 694520 SNDUST. IDL acknowledges support from European Research Council (ERC) Starting Grant 851622 DustOrigin. WRD acknowledges support from STFC Consolidated Grant
ST/S000240/1 to UCL. We also greatly thank both the UCL Widening Participation department and the UKSA Space for All Grants for their support of the ORBYTS programme. 

\section*{Data Availability}

Both GUI and Bayesian versions of the {\sc damocles} code are made publicly available at https://github.com/damocles-code/damocles. Wavelength-calibrated copies of our Gemini GMOS  spectra of SN2012aw are available on the WISeREP archive (https://wiserep.weizmann.ac.il/)

\newpage




\bibliography{example} 

\begin{thebibliography}{93}
\providecommand{\natexlab}[1]{#1}
\providecommand{\url}[1]{\texttt{#1}}
\expandafter\ifx\csname urlstyle\endcsname\relax
  \providecommand{\doi}[1]{doi: #1}\else
  \providecommand{\doi}{doi: \begingroup \urlstyle{rm}\Url}\fi

\bibitem[{Andrews} et~al.(2010){Andrews}, {Gallagher}, {Clayton}, {Sugerman},
  {Chatelain}, {Clem}, {Welch}, {Barlow}, {Ercolano}, {Fabbri}, {Wesson}, and
  {Meixner}]{andrews2010}
J.~E. {Andrews}, J.~S. {Gallagher}, Geoffrey~C. {Clayton}, B.~E.~K. {Sugerman},
  J.~P. {Chatelain}, J.~{Clem}, D.~L. {Welch}, M.~J. {Barlow}, B.~{Ercolano},
  J.~{Fabbri}, R.~{Wesson}, and M.~{Meixner}.
\newblock {SN 2007od: A Type IIP Supernova with Circumstellar Interaction}.
\newblock \emph{ApJ}, 715\penalty0 (1):\penalty0 541--549, May 2010.

\bibitem[{Andrews} and {Smith}(2018)]{Andrews2018}
Jennifer~E. {Andrews} and Nathan {Smith}.
\newblock {Strong late-time circumstellar interaction in the peculiar supernova
  iPTF14hls}.
\newblock \emph{MNRAS}, 477\penalty0 (1):\penalty0 74--79, June 2018.

\bibitem[{Arcavi} et~al.(2017){Arcavi}, {Howell}, {Kasen}, {Bildsten},
  {Hosseinzadeh}, {McCully}, {Wong}, {Katz}, {Gal-Yam}, {Sollerman}, {Taddia},
  {Leloudas}, {Fremling}, {Nugent}, {Horesh}, {Mooley}, {Rumsey}, {Cenko},
  {Graham}, {Perley}, {Nakar}, {Shaviv}, {Bromberg}, {Shen}, {Ofek}, {Cao},
  {Wang}, {Huang}, {Rui}, {Zhang}, {Li}, {Li}, {Zhang}, {Valenti}, {Guevel},
  {Shappee}, {Kochanek}, {Holoien}, {Filippenko}, {Fender}, {Nyholm}, {Yaron},
  {Kasliwal}, {Sullivan}, {Blagorodnova}, {Walters}, {Lunnan}, {Khazov},
  {Andreoni}, {Laher}, {Konidaris}, {Wozniak}, and {Bue}]{arcavi2017}
Iair {Arcavi}, D.~Andrew {Howell}, Daniel {Kasen}, Lars {Bildsten}, Griffin
  {Hosseinzadeh}, Curtis {McCully}, Zheng~Chuen {Wong}, Sarah~Rebekah {Katz},
  Avishay {Gal-Yam}, Jesper {Sollerman}, Francesco {Taddia}, Giorgos
  {Leloudas}, Christoffer {Fremling}, Peter~E. {Nugent}, Assaf {Horesh}, Kunal
  {Mooley}, Clare {Rumsey}, S.~Bradley {Cenko}, Melissa~L. {Graham}, Daniel~A.
  {Perley}, Ehud {Nakar}, Nir~J. {Shaviv}, Omer {Bromberg}, Ken~J. {Shen},
  Eran~O. {Ofek}, Yi~{Cao}, Xiaofeng {Wang}, Fang {Huang}, Liming {Rui},
  Tianmeng {Zhang}, Wenxiong {Li}, Zhitong {Li}, Jujia {Zhang}, Stefano
  {Valenti}, David {Guevel}, Benjamin {Shappee}, Christopher~S. {Kochanek},
  Thomas W.~S. {Holoien}, Alexei~V. {Filippenko}, Rob {Fender}, Anders
  {Nyholm}, Ofer {Yaron}, Mansi~M. {Kasliwal}, Mark {Sullivan}, Nadja
  {Blagorodnova}, Richard~S. {Walters}, Ragnhild {Lunnan}, Danny {Khazov}, Igor
  {Andreoni}, Russ~R. {Laher}, Nick {Konidaris}, Przemek {Wozniak}, and Brian
  {Bue}.
\newblock {Energetic eruptions leading to a peculiar hydrogen-rich explosion of
  a massive star}.
\newblock \emph{\nat}, 551\penalty0 (7679):\penalty0 210--213, November 2017.

\bibitem[Bayless et~al.(2013)Bayless, Pritchard, Roming, Kuin, Brown,
  Botticella, Dall'Ora, Frey, Even, Fryer, Maund, and Fraser]{Bayless2013}
Amanda~J Bayless, Tyler~A. Pritchard, Peter~W.A. Roming, Paul Kuin, Peter~J
  Brown, Maria~Teresa Botticella, Massimo Dall'Ora, Lucille~H Frey, Wesley
  Even, Chris~L Fryer, Justyn~R Maund, and Morgan Fraser.
\newblock {The long-lived UV "plateau" of SN 2012aw}.
\newblock \emph{Astrophysical Journal Letters}, 764\penalty0 (1), 2013.
\newblock ISSN 20418205.

\bibitem[{Bertoldi} et~al.(2003){Bertoldi}, {Carilli}, {Cox}, {Fan}, {Strauss},
  {Beelen}, {Omont}, and {Zylka}]{Bertoldi2003}
F.~{Bertoldi}, C.~L. {Carilli}, P.~{Cox}, X.~{Fan}, M.~A. {Strauss},
  A.~{Beelen}, A.~{Omont}, and R.~{Zylka}.
\newblock {Dust emission from the most distant quasars}.
\newblock \emph{Astronomy and Astrophysics}, 406:\penalty0 L55--L58, July 2003.

\bibitem[Bevan et~al.(2019)Bevan, Wesson, Barlow, {De Looze}, Andrews, Clayton,
  Krafton, Matsuura, and Milisavljevic]{Bevan2019}
A~Bevan, R~Wesson, M~J Barlow, I~{De Looze}, J~E Andrews, G~C Clayton,
  K~Krafton, M~Matsuura, and D~Milisavljevic.
\newblock {A decade of ejecta dust formation in the Type IIn SN 2005ip}.
\newblock \emph{Monthly Notices of the Royal Astronomical Society},
  485\penalty0 (4):\penalty0 5192--5206, 2019.
\newblock ISSN 13652966.

\bibitem[Bevan(2018)]{Bevan2018}
Antonia Bevan.
\newblock {Measuring dust in core-collapse supernovae with a Bayesian approach
  to line profile modelling}.
\newblock \emph{MNRAS}, 480\penalty0 (4):\penalty0 4659--4674, 2018.

\bibitem[Bevan and Barlow(2016)]{Bevan2016}
Antonia Bevan and M.~J. Barlow.
\newblock {Modelling supernova line profile asymmetries to determine ejecta
  dust masses: SN 1987A from days 714 to 3604}.
\newblock \emph{Monthly Notices of the Royal Astronomical Society},
  456\penalty0 (2):\penalty0 1269--1293, 2016.
\newblock ISSN 13652966.

\bibitem[Bevan et~al.(2017)Bevan, Barlow, and Milisavljevic]{Bevan2017}
Antonia Bevan, M.~J. Barlow, and D.~Milisavljevic.
\newblock {Dust masses for SN 1980K, SN1993J and Cassiopeia A from red-blue
  emission line asymmetries}.
\newblock \emph{Monthly Notices of the Royal Astronomical Society},
  465\penalty0 (4):\penalty0 4044--4056, 2017.
\newblock ISSN 13652966.

\bibitem[Bocchio et~al.(2014)Bocchio, Jones, and Slavin]{Bocchio2014}
Marco Bocchio, Anthony~P Jones, and Jonathan~D Slavin.
\newblock {A re-evaluation of dust processing in supernova shock waves}.
\newblock \emph{Astronomy and Astrophysics}, 570:\penalty0 32, 2014.
\newblock ISSN 14320746.

\bibitem[Bose et~al.(2013)Bose, Kumar, Sutaria, Kumar, Roy, Bhatt, Pandey,
  Chandola, Sagar, Misra, and Chakraborti]{Bose2013}
Subhash Bose, Brijesh Kumar, Firoza Sutaria, Brajesh Kumar, Rupak Roy, V.~K.
  Bhatt, S.~B. Pandey, H.~C. Chandola, Ram Sagar, Kuntal Misra, and Sayan
  Chakraborti.
\newblock {Supernova 2012aw - A high-energy clone of archetypal Type IIP SN
  1999em}.
\newblock \emph{Monthly Notices of the Royal Astronomical Society},
  433\penalty0 (3):\penalty0 1871--1891, 2013.
\newblock ISSN 00358711.

\bibitem[{Bouchet} et~al.(1996){Bouchet}, {Danziger}, {Gouiffes}, {della
  Valle}, and {Moneti}]{Bouchet1996}
P.~{Bouchet}, I.~J. {Danziger}, C.~{Gouiffes}, M.~{della Valle}, and
  A.~{Moneti}.
\newblock {SN 1987A: Observations at Later Phases}.
\newblock In Thomas~S. {Kuhn}, editor, \emph{IAU Colloq. 145: Supernovae and
  Supernova Remnants}, page 201, January 1996.

\bibitem[{Chawner} et~al.(2019){Chawner}, {Marsh}, {Matsuura}, {Gomez},
  {Cigan}, {De Looze}, {Barlow}, {Dunne}, {Noriega-Crespo}, and
  {Rho}]{Chawner2019}
H.~{Chawner}, K.~{Marsh}, M.~{Matsuura}, H.~L. {Gomez}, P.~{Cigan}, I.~{De
  Looze}, M.~J. {Barlow}, L.~{Dunne}, A.~{Noriega-Crespo}, and J.~{Rho}.
\newblock {A catalogue of Galactic supernova remnants in the far-infrared:
  revealing ejecta dust in pulsar wind nebulae}.
\newblock \emph{MNRAS}, 483\penalty0 (1):\penalty0 70--118, February 2019.

\bibitem[{Chubb} et~al.(2018){Chubb}, {Naumenko}, {Keely}, {Bartolotto},
  {Macdonald}, {Mukhtar}, {Grachov}, {White}, {Coleman}, {Liu}, {Fazliev},
  {Polovtseva}, {Horneman}, {Campargue}, {Furtenbacher}, {Cs{\'a}sz{\'a}r},
  {Yurchenko}, and {Tennyson}]{Chubb2018b}
Katy~L. {Chubb}, Olga {Naumenko}, Stefan {Keely}, Sebestiano {Bartolotto}, Skye
  {Macdonald}, Mahmoud {Mukhtar}, Andrey {Grachov}, Joe {White}, Eden
  {Coleman}, Anwen {Liu}, Alexander~Z. {Fazliev}, Elena~R. {Polovtseva},
  Veli-Matti {Horneman}, Alain {Campargue}, Tibor {Furtenbacher}, Attila~G.
  {Cs{\'a}sz{\'a}r}, Sergei~N. {Yurchenko}, and Jonathan {Tennyson}.
\newblock {MARVEL analysis of the measured high-resolution rovibrational
  spectra of H$_{2}^{32}S$}.
\newblock \emph{JQSRT}, 218:\penalty0 178--186, October 2018.

\bibitem[{Chugai}(2018)]{Chugai2018}
N.~N. {Chugai}.
\newblock {Extraordinary Supernova iPTF14hls: An Attempt at Interpretation}.
\newblock \emph{Astronomy Letters}, 44\penalty0 (6):\penalty0 370--377, June
  2018.

\bibitem[Dagum and Menon(1998)]{dagum1998}
L.~Dagum and R.~Menon.
\newblock Openmp: an industry standard api for shared-memory programming.
\newblock \emph{IEEE Computational Science and Engineering}, 5\penalty0
  (1):\penalty0 46--55, Jan 1998.
\newblock ISSN 1070-9924.
\newblock \doi{10.1109/99.660313}.

\bibitem[{Darby-Lewis} et~al.(2019){Darby-Lewis}, {Shah}, {Joshi}, {Khan},
  {Kauwo}, {Sethi}, {Bernath}, {Furtenbacher}, {T{\'o}bi{\'a}s},
  {Cs{\'a}sz{\'a}r}, and {Tennyson}]{Darby-Lewis2019}
Daniel {Darby-Lewis}, Het {Shah}, Dhyeya {Joshi}, Fahd {Khan}, Miles {Kauwo},
  Nikhil {Sethi}, Peter~F. {Bernath}, Tibor {Furtenbacher}, Roland
  {T{\'o}bi{\'a}s}, Attila~G. {Cs{\'a}sz{\'a}r}, and Jonathan {Tennyson}.
\newblock {MARVEL analysis of the measured high-resolution spectra of {}1$^{4}$
  NH}.
\newblock \emph{Journal of Molecular Spectroscopy}, 362:\penalty0 69--76,
  August 2019.

\bibitem[{De Looze} et~al.(2017){De Looze}, {Barlow}, {Swinyard}, {Rho},
  {Gomez}, {Matsuura}, and {Wesson}]{DeLooze2017}
I.~{De Looze}, M.~J. {Barlow}, B.~M. {Swinyard}, J.~{Rho}, H.~L. {Gomez},
  M.~{Matsuura}, and R.~{Wesson}.
\newblock {The dust mass in Cassiopeia A from a spatially resolved Herschel
  analysis}.
\newblock \emph{Monthly Notices of the Royal Astronomical Society},
  465\penalty0 (3):\penalty0 3309 (DL2017), March 2017.

\bibitem[{De Looze} et~al.(2019){De Looze}, Barlow, Bandiera, Bevan,
  Bietenholz, Chawner, Gomez, Matsuura, Priestley, and Wesson]{DeLooze2019}
I~{De Looze}, M~J Barlow, R~Bandiera, A~Bevan, M~F Bietenholz, H~Chawner, H~L
  Gomez, M~Matsuura, F~Priestley, and R~Wesson.
\newblock {The dust content of the Crab Nebula}.
\newblock \emph{Monthly Notices of the Royal Astronomical Society},
  488\penalty0 (1):\penalty0 164--182, 2019.
\newblock ISSN 13652966.

\bibitem[{Dessart}(2018)]{Dessart2018}
Luc {Dessart}.
\newblock {A magnetar model for the hydrogen-rich super-luminous supernova
  iPTF14hls}.
\newblock \emph{A\&A}, 610:\penalty0 L10, February 2018.

\bibitem[Dessart et~al.(2021)Dessart, Leonard, Hillier, and
  Pignata]{Dessart2021}
Luc Dessart, Douglas~C Leonard, D~John Hillier, and Giuliano Pignata.
\newblock {Multi-epoch VLT-FORS spectro-polarimetric observations of supernova
  2012aw reveal an asymmetric explosion}.
\newblock \emph{Astronomy and Astrophysics}, page~A8, 2021.

\bibitem[{Draine} and {Lee}(1984)]{draine1984ApJ...285...89D}
B.~T. {Draine} and H.~M. {Lee}.
\newblock {Optical Properties of Interstellar Graphite and Silicate Grains}.
\newblock \emph{The Astrophysical Journal}, 285:\penalty0 89, October 1984.

\bibitem[Dwek et~al.(2007)Dwek, Galliano, and Jones]{Dwek}
Eli Dwek, Frederic Galliano, and Anthony~P Jones.
\newblock {The Evolution of Dust in the Early Universe with Applications to the
  Galaxy SDSS J1148+5251}.
\newblock \emph{The Astrophysical Journal}, 662\penalty0 (2):\penalty0
  927--939, 2007.
\newblock ISSN 0004-637X.

\bibitem[Edwards et~al.(2020)Edwards, Anisman, Changeat, Morvan, Wright, Yip,
  Abdullahi, Ali, Amofa, Antoniou, and et~al.]{Edwards2020}
Billy Edwards, Lara Anisman, Quentin Changeat, Mario Morvan, Sam Wright,
  Kai~Hou Yip, Amiira Abdullahi, Jesmin Ali, Clarry Amofa, Antony Antoniou, and
  et~al.
\newblock Original research by young twinkle students (orbyts): Ephemeris
  refinement of transiting exoplanets ii.
\newblock \emph{Research Notes of the AAS}, 4\penalty0 (7):\penalty0 109, Jul
  2020.
\newblock ISSN 2515-5172.

\bibitem[{Edwards} et~al.(2021{\natexlab{a}}){Edwards}, {Changeat}, {Yip},
  {Tsiaras}, {Taylor}, {Akhtar}, {AlDaghir}, {Bhattarai}, {Bhudia}, {Chapagai},
  {Huang}, {Kabir}, {Khag}, {Khaliq}, {Khatri}, {Kneth}, {Kothari}, {Najmudin},
  {Panchalingam}, {Patel}, {Premachandran}, {Qayyum}, {Rana}, {Shaikh}, {Syed},
  {Theti}, {Zaidani}, {Saraf}, {de Mijolla}, {Caines}, {Kokori}, {Rocchetto},
  {Mallonn}, {Bachschmidt}, {Bosch}, {Bretton}, {Chatelain}, {Deldem}, {Di
  Sisto}, {Evans}, {Fern{\'a}ndez-Laj{\'u}s}, {Guerra}, {Grau Horta}, {Kang},
  {Kim}, {Leroy}, {Lomoz}, {de Haro}, {Hentunen}, {Jongen}, {Molina},
  {Montaigut}, {Naves}, {Raetz}, {Sauer}, {Watkins}, {W{\"u}nsche}, {Zibar},
  {Dunn}, {Tessenyi}, {Savini}, {Tinetti}, and {Tennyson}]{Edwards2021}
Billy {Edwards}, Quentin {Changeat}, Kai~Hou {Yip}, Angelos {Tsiaras}, Jake
  {Taylor}, Bilal {Akhtar}, Josef {AlDaghir}, Pranup {Bhattarai}, Tushar
  {Bhudia}, Aashish {Chapagai}, Michael {Huang}, Danyaal {Kabir}, Vieran
  {Khag}, Summyyah {Khaliq}, Kush {Khatri}, Jaidev {Kneth}, Manisha {Kothari},
  Ibrahim {Najmudin}, Lobanaa {Panchalingam}, Manthan {Patel}, Luxshan
  {Premachandran}, Adam {Qayyum}, Prasen {Rana}, Zain {Shaikh}, Sheryar {Syed},
  Harnam {Theti}, Mahmoud {Zaidani}, Manasvee {Saraf}, Damien {de Mijolla},
  Hamish {Caines}, Anatasia {Kokori}, Marco {Rocchetto}, Matthias {Mallonn},
  Matthieu {Bachschmidt}, Josep~M. {Bosch}, Marc {Bretton}, Philippe
  {Chatelain}, Marc {Deldem}, Romina {Di Sisto}, Phil {Evans}, Eduardo
  {Fern{\'a}ndez-Laj{\'u}s}, Pere {Guerra}, Ferran {Grau Horta}, Wonseok
  {Kang}, Taewoo {Kim}, Arnaud {Leroy}, Franti{\v{s}}ek {Lomoz}, Juan~Lozano
  {de Haro}, Veli-Pekka {Hentunen}, Yves {Jongen}, David {Molina}, Romain
  {Montaigut}, Ramon {Naves}, Manfred {Raetz}, Thomas {Sauer}, Americo
  {Watkins}, Ana{\"e}l {W{\"u}nsche}, Martin {Zibar}, William {Dunn}, Marcell
  {Tessenyi}, Giorgio {Savini}, Giovanna {Tinetti}, and Jonathan {Tennyson}.
\newblock {Original Research by Young Twinkle Students (ORBYTS): ephemeris
  refinement of transiting exoplanets}.
\newblock \emph{MNRAS}, 504\penalty0 (4):\penalty0 5671--5684, July
  2021{\natexlab{a}}.

\bibitem[{Edwards} et~al.(2021{\natexlab{b}}){Edwards}, {Ho}, {Osborne},
  {Deen}, {Hathorn}, {Johnson}, {Patel}, {Vogireddy}, {Waddon}, {Ahmed},
  {Bham}, {Campbell}, {Chummun}, {Crossley}, {Dunsdon}, {Hayes}, {Malik},
  {Marsden}, {Mayfield}, {Mitchell}, {Prosser}, {Rabrenovic}, {Smith},
  {Thomas}, {Kokori}, {Tsiaras}, {Tessenyi}, {Tinetti}, and
  {Tennyson}]{Edwards2021b}
Billy {Edwards}, Cynthia S.~K. {Ho}, Hannah L.~M. {Osborne}, Nabeeha {Deen},
  Ellie {Hathorn}, Solomon {Johnson}, Jiya {Patel}, Varun {Vogireddy}, Ansh
  {Waddon}, Ayuub {Ahmed}, Muhammad {Bham}, Nathan {Campbell}, Zahra {Chummun},
  Nicholas {Crossley}, Reggie {Dunsdon}, Robert {Hayes}, Haroon {Malik}, Frank
  {Marsden}, Lois {Mayfield}, Liston {Mitchell}, Agnes {Prosser}, Valentina
  {Rabrenovic}, Emma {Smith}, Rico {Thomas}, Anastasia {Kokori}, Angelos
  {Tsiaras}, Marcell {Tessenyi}, Giovanna {Tinetti}, and Jonathan {Tennyson}.
\newblock {Original Research By Young Twinkle Students (ORBYTS): Ephemeris
  Refinement of Transiting Exoplanets III}.
\newblock \emph{arXiv e-prints}, art. arXiv:2111.10350, November
  2021{\natexlab{b}}.

\bibitem[Fabbri et~al.(2011)Fabbri, Otsuka, Barlow, Gallagher, Wesson,
  Sugerman, Clayton, Meixner, Andrews, Welch, and Ercolano]{Fabbri2011a}
J~Fabbri, M~Otsuka, M~J Barlow, Joseph~S Gallagher, R~Wesson, B.~E.K. Sugerman,
  Geoffrey~C Clayton, M~Meixner, J~E Andrews, D~L Welch, and B~Ercolano.
\newblock {The effects of dust on the optical and infrared evolution of SN
  2004et}.
\newblock \emph{Monthly Notices of the Royal Astronomical Society},
  418\penalty0 (2):\penalty0 1285--1307, 2011.
\newblock ISSN 00358711.

\bibitem[Fagotti et~al.(2012)Fagotti, Dimai, Quadri, Strabla, Girelli, Quadri,
  Fiorentino, Skvarc, Masi, Fagotti, Dimai, Quadri, Strabla, Girelli, Quadri,
  Fiorentino, Skvarc, and Masi]{Fagotti2012}
P.~Fagotti, A.~Dimai, U.~Quadri, L.~Strabla, R.~Girelli, A.~Quadri,
  L.~Fiorentino, J.~Skvarc, G.~Masi, P.~Fagotti, A.~Dimai, U.~Quadri,
  L.~Strabla, R.~Girelli, A.~Quadri, L.~Fiorentino, J.~Skvarc, and G.~Masi.
\newblock {Supernova 2012aw in M95 = PSN J10435372+1140177.}
\newblock \emph{CBET}, 3054:\penalty0 1, 2012.

\bibitem[{Fesen} and {Weil}(2020)]{Fesen2020}
Robert~A. {Fesen} and Kathryn~E. {Weil}.
\newblock {Detection of Late-time Optical Emission from SN 1941C in NGC 4136}.
\newblock \emph{ApJ}, 890\penalty0 (1):\penalty0 15, February 2020.
\newblock \doi{10.3847/1538-4357/ab67b7}.

\bibitem[Foreman-Mackey et~al.(2013)Foreman-Mackey, Hogg, Lang, and
  Goodman]{Foreman-Mackey2013}
Daniel Foreman-Mackey, David~W Hogg, Dustin Lang, and Jonathan Goodman.
\newblock {emcee : The MCMC Hammer}.
\newblock \emph{Publications of the Astronomical Society of the Pacific},
  125\penalty0 (925):\penalty0 306--312, 2013.
\newblock ISSN 00046280.

\bibitem[Francis et~al.(2020)Francis, Brown, Cameron, Clarke, Dodd, Hurdle,
  Neave, Nowakowska, Patel, Puttock, Redmond, Ruban, Ruban, Savage, Whelan,
  Sidiropoulos, and Muller]{francis2020}
Alistair Francis, Jonathan Brown, Thomas Cameron, Reuben Clarke, Romilly Dodd,
  Jennifer Hurdle, Matthew Neave, Jasmine Nowakowska, Viran Patel, Arianne
  Puttock, Oliver Redmond, Aaron Ruban, Damien Ruban, Meg Savage, Alice Whelan,
  Panagiotis Sidiropoulos, and J.-P Muller.
\newblock A multi-annotator survey of sub-km craters on mars.
\newblock \emph{Data}, 5:\penalty0 70, 08 2020.

\bibitem[Fraser(2016)]{Fraser2016}
Morgan Fraser.
\newblock {The disappearance of the progenitor of SN 2012aw in late-time
  imaging}.
\newblock \emph{Monthly Notices of the Royal Astronomical Society: Letters},
  456\penalty0 (1):\penalty0 L16--L19, 2016.
\newblock ISSN 17453933.

\bibitem[Gall et~al.(2014)Gall, Hjorth, Watson, Dwek, Maund, Fox, Leloudas,
  Malesani, and Day-Jones]{Gall2014}
Christa Gall, Jens Hjorth, Darach Watson, Eli Dwek, Justyn~R Maund, Ori Fox,
  Giorgos Leloudas, Daniele Malesani, and Avril~C Day-Jones.
\newblock {Rapid formation of large dust grains in the luminous supernova
  2010jl}.
\newblock \emph{Nature}, 511\penalty0 (7509):\penalty0 326--329, 2014.
\newblock ISSN 14764687.

\bibitem[{Gallagher} et~al.(2012){Gallagher}, {Sugerman}, {Clayton}, {Andrews},
  {Clem}, {Barlow}, {Ercolano}, {Fabbri}, {Otsuka}, {Wesson}, and
  {Meixner}]{gallagher2012}
Joseph~S. {Gallagher}, B.~E.~K. {Sugerman}, Geoffrey~C. {Clayton}, J.~E.
  {Andrews}, J.~{Clem}, M.~J. {Barlow}, B.~{Ercolano}, J.~{Fabbri},
  M.~{Otsuka}, R.~{Wesson}, and M.~{Meixner}.
\newblock {Optical and Infrared Analysis of Type II SN 2006bc}.
\newblock \emph{ApJ}, 753\penalty0 (2):\penalty0 109, July 2012.

\bibitem[Gelman et~al.(1996)Gelman, Roberts, and Gilks]{Gelman1996}
A.~Gelman, G.~O. Roberts, and W.~R. Gilks.
\newblock Efficient metropolis jumping rules.
\newblock In J.~M. Bernardo, J.~O. Berger, A.~P. Dawid, and A.~F.~M. Smith,
  editors, \emph{Bayesian Statistics}, pages 599--608. Oxford University Press,
  Oxford, 1996.

\bibitem[Gerasimovic(1933)]{Gerasimovic1933a}
B.P. Gerasimovic.
\newblock {The contours of emission lines in expanding nebular envelopes}.
\newblock \emph{Zeitschrift f{\"{u}}r Astrophysik}, 7:\penalty0 335, 1933.

\bibitem[Gomez et~al.(2012)Gomez, Krause, Barlow, Swinyard, Owen, Clark,
  Matsuura, Gomez, Rho, Besel, Bouwman, Gear, Henning, Ivison, Polehampton, and
  Sibthorpe]{Gomez2012}
H.~L Gomez, O~Krause, M~J Barlow, B~M Swinyard, P~J Owen, C.~J.R. Clark,
  M~Matsuura, E~L Gomez, J~Rho, M.~A. Besel, J~Bouwman, W~K Gear, Th~Henning,
  R~J Ivison, E~T Polehampton, and B~Sibthorpe.
\newblock {A cool dust factory in the Crab Nebula: A Herschel study of the
  filaments}.
\newblock \emph{Astrophysical Journal}, 760\penalty0 (1):\penalty0 96, 2012.
\newblock ISSN 15384357.

\bibitem[{Goodman} and {Weare}(2010)]{Goodman2010}
Jonathan {Goodman} and Jonathan {Weare}.
\newblock {Ensemble samplers with affine invariance}.
\newblock \emph{Communications in Applied Mathematics and Computational
  Science}, 5\penalty0 (1):\penalty0 65--80, January 2010.

\bibitem[{Grafton-Waters} et~al.(2021){Grafton-Waters}, {Ahmed}, {Henson},
  {Hinds-Williams}, {Ivanova}, {Marshall}, {Udueni}, {Theodorakis}, and
  {Dunn}]{Grafton-Waters2021}
S.~{Grafton-Waters}, M.~{Ahmed}, S.~{Henson}, F.~{Hinds-Williams},
  B.~{Ivanova}, E.~{Marshall}, H.~{Udueni}, D.~{Theodorakis}, and W.~{Dunn}.
\newblock {A Study of the Soft X-Ray Emission Lines in NGC 4151. I. Kinematic
  Properties of the Plasma Wind}.
\newblock \emph{Research Notes of the American Astronomical Society},
  5\penalty0 (7):\penalty0 172, July 2021.

\bibitem[{Holdship} et~al.(2019){Holdship}, {Viti}, {Codella}, {Rawlings},
  {Jimenez-Serra}, {Ayalew}, {Curtis}, {Habib}, {Lawrence}, {Warsame}, and
  {Horn}]{Holdship2019}
Jonathan {Holdship}, Serena {Viti}, Claudio {Codella}, Jonathan {Rawlings},
  Izaskun {Jimenez-Serra}, Yenabeb {Ayalew}, Justin {Curtis}, Annur {Habib},
  Jamel {Lawrence}, Sumaya {Warsame}, and Sarah {Horn}.
\newblock {Observations of CH$_{3}$OH and CH$_{3}$CHO in a Sample of
  Protostellar Outflow Sources}.
\newblock \emph{ApJ}, 880\penalty0 (2):\penalty0 138, August 2019.

\bibitem[Indebetouw et~al.(2014)Indebetouw, Matsuura, Dwek, Zanardo, Barlow,
  Baes, Bouchet, Burrows, Chevalier, Clayton, Fransson, Gaensler, Kirshner,
  Laki{\'{c}}evi{\'{c}}, Long, Lundqvist, Mart{\'{i}}-Vidal, Marcaide, McCray,
  Meixner, Ng, Park, Sonneborn, Staveley-Smith, Vlahakis, and {Van
  Loon}]{Indebetouw2014}
R~Indebetouw, M~Matsuura, E~Dwek, G~Zanardo, M~J Barlow, M~Baes, P~Bouchet, D~N
  Burrows, R~Chevalier, G~C Clayton, C~Fransson, B.~Gaensler, R.~Kirshner,
  M.~Laki{\'{c}}evi{\'{c}}, K.~S. Long, P.~Lundqvist, I.~Mart{\'{i}}-Vidal,
  J.~Marcaide, R.~McCray, M.~Meixner, C.~Y. Ng, S.~Park, G.~Sonneborn,
  L.~Staveley-Smith, C~Vlahakis, and J~{Van Loon}.
\newblock {Dust production and particle acceleration in supernova 1987a
  revealed with Alma}.
\newblock \emph{Astrophysical Journal Letters}, 782\penalty0 (1):\penalty0 11,
  2014.
\newblock ISSN 20418205.

\bibitem[Jerkstrand et~al.(2014)Jerkstrand, Smartt, Fraser, Fransson,
  Sollerman, Taddia, and Kotak]{Jerkstrand2014}
A~Jerkstrand, S~J Smartt, M~Fraser, C~Fransson, J~Sollerman, F~Taddia, and
  R~Kotak.
\newblock {The nebular spectra of SN 2012aw and constraints on stellar
  nucleosynthesis from oxygen emission lines}.
\newblock \emph{Monthly Notices of the Royal Astronomical Society},
  439\penalty0 (4):\penalty0 3694--3703, 2014.
\newblock ISSN 13652966.

\bibitem[Kirchschlager and Bertrang(2020)]{Kirchschlager2020}
Florian Kirchschlager and Gesa H.-M. Bertrang.
\newblock {Self-scattering of non-spherical dust grains}.
\newblock \emph{Astronomy {\&} Astrophysics}, 638:\penalty0 A116, jun 2020.
\newblock ISSN 0004-6361.

\bibitem[Kirchschlager et~al.(2019)Kirchschlager, Schmidt, Barlow, Fogerty,
  Bevan, and Priestley]{Kirchschlager2019}
Florian Kirchschlager, Franziska~D Schmidt, M~J Barlow, Erica~L Fogerty,
  Antonia Bevan, and Felix~D Priestley.
\newblock {Dust survival rates in clumps passing through the Cas A reverse
  shock – I. Results for a range of clump densities}.
\newblock \emph{Monthly Notices of the Royal Astronomical Society},
  489\penalty0 (4):\penalty0 4465--4496, 2019.
\newblock ISSN 13652966.

\bibitem[{Kirchschlager} et~al.(2022){Kirchschlager}, {Schmidt}, {Barlow}, {De
  Looze}, and {Sartorio}]{Kirchschlager2022}
Florian {Kirchschlager}, Franziska~D. {Schmidt}, M.~J. {Barlow}, Ilse {De
  Looze}, and Nina~S. {Sartorio}.
\newblock {Dust survival rates in clumps passing through the CasA reverse shock
  -- II. The impact of magnetic fields}.
\newblock \emph{arXiv e-prints}, art. arXiv:2210.06763, October 2022.

\bibitem[Kotak et~al.(2009)Kotak, Meikle, Farrah, Gerardy, Foley, {Van Dyk},
  Fransson, Lundqvist, Sollerman, Fesen, Filippenko, Mattila, Silverman,
  Andersen, H{\"{o}}flich, Pozzo, and Wheeler]{Kotak2009}
R~Kotak, W.~P.S. Meikle, D~Farrah, C~L Gerardy, R~J Foley, S~D {Van Dyk},
  C~Fransson, P~Lundqvist, J~Sollerman, R~Fesen, A~V Filippenko, S~Mattila, J~M
  Silverman, A~C Andersen, P~A H{\"{o}}flich, M~Pozzo, and J~C Wheeler.
\newblock {Dust and the type II-plateau supernova 2004et}.
\newblock \emph{Astrophysical Journal}, 704\penalty0 (1):\penalty0 306--323,
  2009.
\newblock ISSN 15384357.

\bibitem[{Kozma} and {Fransson}(1998)]{Kozma1998b}
Cecilia {Kozma} and Claes {Fransson}.
\newblock {Late Spectral Evolution of SN 1987A. II. Line Emission}.
\newblock \emph{ApJ}, 497\penalty0 (1):\penalty0 431--457, April 1998.

\bibitem[{Laporte} et~al.(2017){Laporte}, {Ellis}, {Boone}, {Bauer},
  {Qu{\'e}nard}, {Roberts-Borsani}, {Pell{\'o}}, {P{\'e}rez-Fournon}, and
  {Streblyanska}]{Laporte2017b}
N.~{Laporte}, R.~S. {Ellis}, F.~{Boone}, F.~E. {Bauer}, D.~{Qu{\'e}nard}, G.~W.
  {Roberts-Borsani}, R.~{Pell{\'o}}, I.~{P{\'e}rez-Fournon}, and
  A.~{Streblyanska}.
\newblock {Dust in the Reionization Era: ALMA Observations of a z = 8.38
  Gravitationally Lensed Galaxy}.
\newblock \emph{The Astrophysical Journal Letters}, 837\penalty0 (2):\penalty0
  L21, March 2017.

\bibitem[{Li} et~al.(2015){Li}, {Wang}, and {Zhang}]{Li2015}
Wenxiong {Li}, Xiaofeng {Wang}, and Tianmeng {Zhang}.
\newblock {Spectroscopic Classification of CSS141118:092034+504148 as a Type
  II-P Supernova}.
\newblock \emph{The Astronomer's Telegram}, 6898:\penalty0 1, January 2015.

\bibitem[{Lucy} et~al.(1989){Lucy}, {Danziger}, {Gouiffes}, and
  {Bouchet}]{Lucy1989}
L.~B. {Lucy}, I.~J. {Danziger}, C.~{Gouiffes}, and P.~{Bouchet}.
\newblock \emph{{Dust Condensation in the Ejecta of SN 1987A}}, volume 350,
  page 164.
\newblock Springer-Verlag, 1989.

\bibitem[{Lucy} et~al.(1991){Lucy}, {Danziger}, {Gouiffes}, and
  {Bouchet}]{Lucy1991}
L.~B. {Lucy}, I.~J. {Danziger}, C.~{Gouiffes}, and P.~{Bouchet}.
\newblock {Dust Condensation in the Ejecta of Supernova 1987A - Part Two}.
\newblock In Stanford~E. {Woosley}, editor, \emph{Supernovae}, page~82, January
  1991.

\bibitem[Matsuura et~al.(2011)Matsuura, Dwek, Meixner, Otsuka, Babler, Barlow,
  Roman-Duval, Engelbracht, Sandstrom, Laki{\'{c}}evi{\'{c}}, {Van Loon},
  Sonneborn, Clayton, Long, Lundqvist, Nozawa, Gordon, Hony, Panuzzo, Okumura,
  Misselt, Montiel, and Sauvage]{Matsuura}
M~Matsuura, E~Dwek, M~Meixner, M~Otsuka, B~Babler, M~J Barlow, J~Roman-Duval,
  C~Engelbracht, K~Sandstrom, M~Laki{\'{c}}evi{\'{c}}, J~Th {Van Loon},
  G~Sonneborn, G~C Clayton, K~S Long, P~Lundqvist, T~Nozawa, K~D Gordon,
  S~Hony, P~Panuzzo, K~Okumura, K~A Misselt, E~Montiel, and M~Sauvage.
\newblock {Herschel detects a massive dust reservoir in supernova 1987A}.
\newblock \emph{Science}, 333\penalty0 (6047):\penalty0 1258--1261, 2011.
\newblock ISSN 10959203.

\bibitem[{Matsuura} et~al.(2015){Matsuura}, {Dwek}, {Barlow}, {Babler}, {Baes},
  {Meixner}, {Cernicharo}, {Clayton}, {Dunne}, {Fransson}, {Fritz}, {Gear},
  {Gomez}, {Groenewegen}, {Indebetouw}, {Ivison}, {Jerkstrand}, {Lebouteiller},
  {Lim}, {Lundqvist}, {Pearson}, {Roman-Duval}, {Royer}, {Staveley-Smith},
  {Swinyard}, {van Hoof}, {van Loon}, {Verstappen}, {Wesson}, {Zanardo},
  {Blommaert}, {Decin}, {Reach}, {Sonneborn}, {Van de Steene}, and
  {Yates}]{Matsuura2015}
M.~{Matsuura}, E.~{Dwek}, M.~J. {Barlow}, B.~{Babler}, M.~{Baes}, M.~{Meixner},
  Jos{\'e} {Cernicharo}, Geoff~C. {Clayton}, L.~{Dunne}, C.~{Fransson}, Jacopo
  {Fritz}, Walter {Gear}, H.~L. {Gomez}, M.~A.~T. {Groenewegen},
  R.~{Indebetouw}, R.~J. {Ivison}, A.~{Jerkstrand}, V.~{Lebouteiller}, T.~L.
  {Lim}, P.~{Lundqvist}, C.~P. {Pearson}, J.~{Roman-Duval}, P.~{Royer}, Lister
  {Staveley-Smith}, B.~M. {Swinyard}, P.~A.~M. {van Hoof}, J.~Th. {van Loon},
  Joris {Verstappen}, Roger {Wesson}, Giovanna {Zanardo}, Joris A.~D.~L.
  {Blommaert}, Leen {Decin}, W.~T. {Reach}, George {Sonneborn}, Griet~C. {Van
  de Steene}, and Jeremy~A. {Yates}.
\newblock {A Stubbornly Large Mass of Cold Dust in the Ejecta of Supernova
  1987A}.
\newblock \emph{ApJ}, 800\penalty0 (1):\penalty0 50, February 2015.

\bibitem[Mauerhan and Smith(2012)]{Mauerhan2012}
Jon Mauerhan and Nathan Smith.
\newblock {Supernova 1998S at 14 years postmortem: Continuing circumstellar
  interaction and dust formation}.
\newblock \emph{Monthly Notices of the Royal Astronomical Society},
  424\penalty0 (4):\penalty0 2659--2666, 2012.
\newblock ISSN 00358711.

\bibitem[{McKemmish} et~al.(2017{\natexlab{a}}){McKemmish}, {Chubb}, {Rivlin},
  {Baker}, {Gorman}, {Heward}, {Dunn}, and {Tessenyi}]{McKemmish2017b}
Laura~K. {McKemmish}, Katy~L. {Chubb}, Tom {Rivlin}, Jack~S. {Baker}, Maire~N.
  {Gorman}, Anita {Heward}, William {Dunn}, and Marcell {Tessenyi}.
\newblock {Bringing pupils into the ORBYTS of research}.
\newblock \emph{Astronomy and Geophysics}, 58\penalty0 (5):\penalty0
  5.11--5.11, October 2017{\natexlab{a}}.

\bibitem[{McKemmish} et~al.(2017{\natexlab{b}}){McKemmish}, {Masseron},
  {Sheppard}, {Sandeman}, {Schofield}, {Furtenbacher}, {Cs{\'a}sz{\'a}r},
  {Tennyson}, and {Sousa-Silva}]{McKemmish2017a}
Laura~K. {McKemmish}, Thomas {Masseron}, Samuel {Sheppard}, Elizabeth
  {Sandeman}, Zak {Schofield}, Tibor {Furtenbacher}, Attila~G.
  {Cs{\'a}sz{\'a}r}, Jonathan {Tennyson}, and Clara {Sousa-Silva}.
\newblock {Marvel Analysis of the Measured High-resolution Rovibronic Spectra
  of TiO}.
\newblock \emph{ApJs}, 228\penalty0 (2):\penalty0 15, February
  2017{\natexlab{b}}.

\bibitem[{McKemmish} et~al.(2018){McKemmish}, {Borsovszky}, {Goodhew},
  {Sheppard}, {Bennett}, {Martin}, {Singh}, {Sturgeon}, {Furtenbacher},
  {Cs{\'a}sz{\'a}r}, and {Tennyson}]{McKemmish2018}
Laura~K. {McKemmish}, Jasmin {Borsovszky}, Katie~L. {Goodhew}, Samuel
  {Sheppard}, Aphra F.~V. {Bennett}, Alfie D.~J. {Martin}, Amrik {Singh},
  Callum A.~J. {Sturgeon}, Tibor {Furtenbacher}, Attila~G. {Cs{\'a}sz{\'a}r},
  and Jonathan {Tennyson}.
\newblock {MARVEL Analysis of the Measured High-resolution Rovibronic Spectra
  of $^{90}$Zr$^{16}$O}.
\newblock \emph{ApJ}, 867\penalty0 (1):\penalty0 33, November 2018.

\bibitem[Micelotta et~al.(2016)Micelotta, Dwek, and Slavin]{Micelotta2016}
Elisabetta~R Micelotta, Eli Dwek, and Jonathan~D Slavin.
\newblock {Dust destruction by the reverse shock in the Cassiopeia A supernova
  remnant}.
\newblock \emph{Astronomy and Astrophysics}, 590:\penalty0 65, 2016.
\newblock ISSN 14320746.

\bibitem[Milisavljevic et~al.(2012)Milisavljevic, Fesen, Chevalier, Kirshner,
  Challis, and Turatto]{Milisavljevic2012}
Dan Milisavljevic, Robert~A. Fesen, Roger~A. Chevalier, Robert~P. Kirshner,
  Peter Challis, and Massimo Turatto.
\newblock {Late-time optical emission from core-collapse supernovae}.
\newblock \emph{Astrophysical Journal}, 751\penalty0 (1):\penalty0 25, may
  2012.
\newblock ISSN 15384357.

\bibitem[{Morgan} and {Edmunds}(2003)]{Morgan2003a}
H.~L. {Morgan} and M.~G. {Edmunds}.
\newblock {Dust formation in early galaxies}.
\newblock \emph{Monthly Notices of the Royal Astronomical Society},
  343\penalty0 (2):\penalty0 427--442, August 2003.

\bibitem[{Moriya} et~al.(2020){Moriya}, {Mazzali}, and {Pian}]{Moriya2020}
Takashi~J. {Moriya}, Paolo~A. {Mazzali}, and Elena {Pian}.
\newblock {iPTF14hls as a variable hyper-wind from a very massive star}.
\newblock \emph{MNRAS}, 491\penalty0 (1):\penalty0 1384--1390, January 2020.

\bibitem[Nath et~al.(2008)Nath, Laskar, and Shull]{Nath}
Biman~B Nath, Tanmoy Laskar, and J~Michael Shull.
\newblock {Dust Sputtering by Reverse Shocks in Supernova Remnants}.
\newblock \emph{The Astrophysical Journal}, 682\penalty0 (2):\penalty0
  1055--1064, 2008.
\newblock ISSN 0004-637X.

\bibitem[Niculescu-Duvaz et~al.(2021)Niculescu-Duvaz, Barlow, Bevan,
  Milisavljevic, and {De Looze}]{Niculescu-Duvaz2021}
Maria Niculescu-Duvaz, Michael~J Barlow, Antonia Bevan, Danny Milisavljevic,
  and Ilse {De Looze}.
\newblock {The dust mass in Cassiopeia A from infrared and optical line flux
  differences}.
\newblock \emph{Monthly Notices of the Royal Astronomical Society}, 2021.
\newblock ISSN 0035-8711.

\bibitem[{Niculescu-Duvaz} et~al.(2022){Niculescu-Duvaz}, {Barlow}, {Bevan},
  {Wesson}, {Milisavljevic}, {De Looze}, {Clayton}, {Krafton}, {Matsuura}, and
  {Brady}]{Niculescu-Duvaz2022}
Maria {Niculescu-Duvaz}, Michael~J {Barlow}, Antonia {Bevan}, Roger {Wesson},
  Danny {Milisavljevic}, Ilse {De Looze}, Geoff~C. {Clayton}, Kelsie {Krafton},
  Mikako {Matsuura}, and Ryan {Brady}.
\newblock {Dust masses for a large sample of core-collapse supernovae from
  optical emission line asymmetries: dust formation on 30-year timescales}.
\newblock \emph{arXiv e-prints}, art. arXiv:2204.14179, April 2022.

\bibitem[Nozawa et~al.(2003)Nozawa, Kozasa, Umeda, Maeda, and
  Nomoto]{Nozawa2003}
Takaya Nozawa, Takashi Kozasa, Hideyuki Umeda, Keiichi Maeda, and Ken'ichi
  Nomoto.
\newblock {Dust in the Early Universe: Dust Formation in the Ejecta of
  Population III Supernovae}.
\newblock \emph{The Astrophysical Journal}, 598\penalty0 (2):\penalty0
  785--803, 2003.
\newblock ISSN 0004-637X.

\bibitem[{Owen} and {Barlow}(2015)]{owen2015}
P.~J. {Owen} and M.~J. {Barlow}.
\newblock {The Dust and Gas Content of the Crab Nebula}.
\newblock \emph{ApJ}, 801\penalty0 (2):\penalty0 141, March 2015.
\newblock \doi{10.1088/0004-637X/801/2/141}.

\bibitem[{Poznanski} et~al.(2012){Poznanski}, {Nugent}, {Ofek}, {Gal-Yam}, and
  {Kasliwal}]{poznanski2012}
Dovi {Poznanski}, Peter~E. {Nugent}, Eran~O. {Ofek}, Avishay {Gal-Yam}, and
  Mansi~M. {Kasliwal}.
\newblock {PTF observations of SN2012aw (PTF12bvh) and explosion date
  constraints}.
\newblock \emph{The Astronomer's Telegram}, 3996:\penalty0 1, March 2012.

\bibitem[{Priestley} et~al.(2019){Priestley}, {Barlow}, and {De
  Looze}]{Priestley2019}
F.~D. {Priestley}, M.~J. {Barlow}, and I.~{De Looze}.
\newblock {The mass, location, and heating of the dust in the Cassiopeia A
  supernova remnant}.
\newblock \emph{MNRAS}, 485\penalty0 (1):\penalty0 440--451, May 2019.

\bibitem[{Priestley} et~al.(2020){Priestley}, {Barlow}, {De Looze}, and
  {Chawner}]{priestley2020}
F.~D. {Priestley}, M.~J. {Barlow}, I.~{De Looze}, and H.~{Chawner}.
\newblock {Dust masses and grain size distributions of a sample of Galactic
  pulsar wind nebulae}.
\newblock \emph{MNRAS}, 491\penalty0 (4):\penalty0 6020--6031, February 2020.
\newblock \doi{10.1093/mnras/stz3434}.

\bibitem[Priestley et~al.(2021)Priestley, Chawner, Matsuura, {De Looze},
  Barlow, and Gomez]{Priestley2021}
F.~D. Priestley, H.~Chawner, M.~Matsuura, I.~{De Looze}, M.~J. Barlow, and
  H.~L. Gomez.
\newblock {Revisiting the dust destruction efficiency of supernovae}.
\newblock \emph{Monthly Notices of the Royal Astronomical Society},
  500\penalty0 (2):\penalty0 2543--2553, jan 2021.
\newblock ISSN 13652966.

\bibitem[{Prokhorov} et~al.(2021){Prokhorov}, {Moraghan}, and
  {Vink}]{Prokhorov2021}
D.~A. {Prokhorov}, A.~{Moraghan}, and J.~{Vink}.
\newblock {Search for gamma rays from SNe with a variable-size
  sliding-time-window analysis of the Fermi-LAT data}.
\newblock \emph{MNRAS}, 505\penalty0 (1):\penalty0 1413--1421, July 2021.

\bibitem[Rho et~al.(2009)Rho, Reach, Tappe, Rudnick, Kozasa, Hwang, Andersen,
  Gomez, DeLaney, Dunne, and Slavin]{Rho2009}
J.~Rho, W.~T. Reach, A.~Tappe, L.~Rudnick, T.~Kozasa, U.~Hwang, M.~Andersen,
  H.~Gomez, T.~DeLaney, L.~Dunne, and J.~Slavin.
\newblock {Dust Formation Observed in Young Supernova Remnants with Spitzer}.
\newblock \emph{Astronomical Society of the Pacific Conference Series},
  414:\penalty0 22, 2009.

\bibitem[Sarangi and Cherchneff(2015)]{Sarangi2015}
Arkaprabha Sarangi and Isabelle Cherchneff.
\newblock {Condensation of dust in the ejecta of Type II-P supernovae}.
\newblock \emph{Astronomy and Astrophysics}, 575:\penalty0 95, 2015.
\newblock ISSN 14320746.

\bibitem[Silvia et~al.(2010)Silvia, Smith, and Shull]{Silvia2010}
Devin~W Silvia, Britton~D Smith, and J.~M. Shull.
\newblock {Numerical simulations of supernova dust destruction. I.
  Cloud-crushing and post-processed grain sputtering}.
\newblock \emph{Astrophysical Journal}, 715\penalty0 (2):\penalty0 1575--1590,
  2010.
\newblock ISSN 15384357.

\bibitem[Simon et~al.(2020)Simon, Mallaburn, and Seton]{simon2020}
S.~Simon, Andrea Mallaburn, and Linda Seton.
\newblock \emph{Chapter 10. Enhancing School Students’ Engagement in
  Chemistry Through a University-led Enrichment Programme}, pages 192--224.
\newblock 07 2020.
\newblock ISBN 978-1-78801-508-0.

\bibitem[{Slavin} et~al.(2020){Slavin}, {Dwek}, {Mac Low}, and
  {Hill}]{Slavin2020}
Jonathan~D. {Slavin}, Eli {Dwek}, Mordecai-Mark {Mac Low}, and Alex~S. {Hill}.
\newblock {The Dynamics, Destruction, and Survival of Supernova-formed Dust
  Grains}.
\newblock \emph{The Astrophysical Journal}, 902\penalty0 (2):\penalty0 135,
  October 2020.

\bibitem[Smith et~al.(2008)Smith, Foley, and Filippenko]{Smith2008}
Nathan Smith, Ryan~J Foley, and Alexei~V Filippenko.
\newblock {Dust Formation and He ii $\lambda$4686 Emission in the Dense Shell
  of the Peculiar Type Ib Supernova 2006jc}.
\newblock \emph{The Astrophysical Journal}, 680\penalty0 (1):\penalty0
  568--579, 2008.
\newblock ISSN 0004-637X.

\bibitem[{Soker} and {Gilkis}(2018)]{soker2018}
Noam {Soker} and Avishai {Gilkis}.
\newblock {Explaining iPTF14hls as a common-envelope jets supernova}.
\newblock \emph{MNRAS}, 475\penalty0 (1):\penalty0 1198--1202, March 2018.

\bibitem[{Sollerman} et~al.(2019){Sollerman}, {Taddia}, {Arcavi}, {Fremling},
  {Fransson}, {Burke}, {Cenko}, {Andersen}, {Andreoni}, {Barbarino},
  {Blagorodova}, {Brink}, {Filippenko}, {Gal-Yam}, {Hiramatsu}, {Hosseinzadeh},
  {Howell}, {de Jaeger}, {Lunnan}, {McCully}, {Perley}, {Tartaglia},
  {Terreran}, {Valenti}, and {Wang}]{sollerman2019}
J.~{Sollerman}, F.~{Taddia}, I.~{Arcavi}, C.~{Fremling}, C.~{Fransson},
  J.~{Burke}, S.~B. {Cenko}, O.~{Andersen}, I.~{Andreoni}, C.~{Barbarino},
  N.~{Blagorodova}, T.~G. {Brink}, A.~V. {Filippenko}, A.~{Gal-Yam},
  D.~{Hiramatsu}, G.~{Hosseinzadeh}, D.~A. {Howell}, T.~{de Jaeger},
  R.~{Lunnan}, C.~{McCully}, D.~A. {Perley}, L.~{Tartaglia}, G.~{Terreran},
  S.~{Valenti}, and X.~{Wang}.
\newblock {Late-time observations of the extraordinary Type II supernova
  iPTF14hls}.
\newblock \emph{A\&A}, 621:\penalty0 A30, January 2019.

\bibitem[{Sousa-Silva} et~al.(2018){Sousa-Silva}, {McKemmish}, {Chubb},
  {Gorman}, {Baker}, {Barton}, {Rivlin}, and {Tennyson}]{Sousa2018}
Clara {Sousa-Silva}, Laura~K. {McKemmish}, Katy~L. {Chubb}, Maire~N. {Gorman},
  Jack~S. {Baker}, Emma~J. {Barton}, Tom {Rivlin}, and Jonathan {Tennyson}.
\newblock {Original Research By Young Twinkle Students (ORBYTS): when can
  students start performing original research?}
\newblock \emph{Physics Education}, 53\penalty0 (1):\penalty0 015020, January
  2018.

\bibitem[Springob et~al.(2005)Springob, Haynes, Giovanelli, and
  Kent]{Springob2005}
Christopher~M Springob, Martha~P Haynes, Riccardo Giovanelli, and Brian~R Kent.
\newblock {A Digital Archive of H i 21 Centimeter Line Spectra of Optically
  Targeted Galaxies}.
\newblock \emph{The Astrophysical Journal Supplement Series}, 160\penalty0
  (1):\penalty0 149--162, 2005.
\newblock ISSN 0067-0049.

\bibitem[Sugerman et~al.(2006)Sugerman, Ercolano, Barlow, Tielens, Clayton,
  Zijlstra, Meixner, Speck, Gledhill, Panagia, Cohen, Gordon, Meyer, Fabbri,
  Bowey, Welch, Regan, and Kennicutt]{Sugerman2006}
Ben~E.K. Sugerman, Barbara Ercolano, M~J Barlow, A.~G.G.M. Tielens, Geoffrey~C
  Clayton, Albert~A Zijlstra, Margaret Meixner, Angela Speck, Tim~M Gledhill,
  Nino Panagia, Martin Cohen, Karl~D Gordon, Martin Meyer, Joanna Fabbri,
  Janet~E Bowey, Douglas~L Welch, Michael~W Regan, and Robert~C Kennicutt.
\newblock {Massive-star supernovae as major dust factories}.
\newblock \emph{Science}, 313\penalty0 (5784):\penalty0 196--200, 2006.
\newblock ISSN 00368075.

\bibitem[{Szalai} and {Vink{\'o}}(2013)]{szalai2013}
T.~{Szalai} and J.~{Vink{\'o}}.
\newblock {Twelve type II-P supernovae seen with the eyes of Spitzer}.
\newblock \emph{A\&A}, 549:\penalty0 A79, January 2013.

\bibitem[{Szalai} et~al.(2011){Szalai}, {Vink{\'o}}, {Balog}, {G{\'a}sp{\'a}r},
  {Block}, and {Kiss}]{szalai2011}
T.~{Szalai}, J.~{Vink{\'o}}, Z.~{Balog}, A.~{G{\'a}sp{\'a}r}, M.~{Block}, and
  L.~L. {Kiss}.
\newblock {Dust formation in the ejecta of the type II-P supernova 2004dj}.
\newblock \emph{A\&A}, 527:\penalty0 A61, March 2011.

\bibitem[{Temim} et~al.(2017){Temim}, {Dwek}, {Arendt}, {Borkowski},
  {Reynolds}, {Slane}, {Gelfand}, and {Raymond}]{Temim2017}
Tea {Temim}, Eli {Dwek}, Richard~G. {Arendt}, Kazimierz~J. {Borkowski},
  Stephen~P. {Reynolds}, Patrick {Slane}, Joseph~D. {Gelfand}, and John~C.
  {Raymond}.
\newblock {A Massive Shell of Supernova-formed Dust in SNR G54.1+0.3}.
\newblock \emph{The Astrophysical Journal}, 836\penalty0 (1):\penalty0 129,
  February 2017.

\bibitem[{Uno} and {Maeda}(2020)]{Uno2020}
Kohki {Uno} and Keiichi {Maeda}.
\newblock {A Wind-driven Model: Application to Peculiar Transients AT2018cow
  and iPTF14hls}.
\newblock \emph{ApJ}, 897\penalty0 (2):\penalty0 156, July 2020.

\bibitem[{van Dokkum}(2001)]{vanDokkum2001}
Pieter~G. {van Dokkum}.
\newblock {Cosmic-Ray Rejection by Laplacian Edge Detection}.
\newblock \emph{PASP}, 113\penalty0 (789):\penalty0 1420--1427, November 2001.

\bibitem[{Van Dyk} et~al.(2012){Van Dyk}, Cenko, Poznanski, Arcavi, Gal-Yam,
  Filippenko, Silverio, Stockton, Cuillandre, Marcy, Howard, and
  Isaacson]{VanDyk2012}
Schuyler~D {Van Dyk}, S~Bradley Cenko, Dovi Poznanski, Iair Arcavi, Avishay
  Gal-Yam, Alexei~V Filippenko, Kathryn Silverio, Alan Stockton, Jean~Charles
  Cuillandre, Geoffrey~W Marcy, Andrew~W Howard, and Howard Isaacson.
\newblock {The red supergiant progenitor of supernova 2012aw (PTF12bvh) in
  messier 95}.
\newblock \emph{Astrophysical Journal}, 756\penalty0 (2), 2012.
\newblock ISSN 15384357.

\bibitem[Watson et~al.(2015)Watson, Christensen, Knudsen, Richard, Gallazzi,
  and Micha{\l}owski]{Watson2015a}
Darach Watson, Lise Christensen, Kirsten~Kraiberg Knudsen, Johan Richard, Anna
  Gallazzi, and Micha{\l}~Jerzy Micha{\l}owski.
\newblock {A dusty, normal galaxy in the epoch of reionization}.
\newblock \emph{Nature}, 519\penalty0 (7543):\penalty0 327--330, 2015.

\bibitem[Wesson et~al.(2015)Wesson, Barlow, Matsuura, and Ercolano]{Wesson2015}
R~Wesson, M~J Barlow, M~Matsuura, and B~Ercolano.
\newblock {The timing and location of dust formation in the remnant of SN
  1987A}.
\newblock \emph{Monthly Notices of the Royal Astronomical Society},
  446\penalty0 (2):\penalty0 2089--2101, 2015.
\newblock ISSN 13652966.

\bibitem[{Wibisono} et~al.(2020){Wibisono}, {Branduardi-Raymont}, {Dunn},
  {Coates}, {Weigt}, {Jackman}, {Yao}, {Tao}, {Allegrini}, {Grodent},
  {Chatterton}, {Gerasimova}, {Kloss}, {Milovi{\'c}}, {Orlandiayni}, {Preidl},
  {Radler}, {Summhammer}, and {Fleming}]{wibisono2020}
A.~D. {Wibisono}, G.~{Branduardi-Raymont}, W.~R. {Dunn}, A.~J. {Coates}, D.~M.
  {Weigt}, C.~M. {Jackman}, Z.~H. {Yao}, C.~{Tao}, F.~{Allegrini},
  D.~{Grodent}, J.~{Chatterton}, A.~{Gerasimova}, L.~{Kloss}, J.~{Milovi{\'c}},
  L.~{Orlandiayni}, A.~K. {Preidl}, C.~{Radler}, L.~{Summhammer}, and
  D.~{Fleming}.
\newblock {Temporal and Spectral Studies by XMM-Newton of Jupiter's X-ray
  Auroras During a Compression Event}.
\newblock \emph{Journal of Geophysical Research (Space Physics)}, 125\penalty0
  (5):\penalty0 e27676, May 2020.

\bibitem[{Yuan} et~al.(2018){Yuan}, {Liao}, {Xin}, {Li}, {Fan}, {Zhang}, {Hu},
  and {Bi}]{Yuan2018}
Qiang {Yuan}, Neng-Hui {Liao}, Yu-Liang {Xin}, Ye~{Li}, Yi-Zhong {Fan}, Bing
  {Zhang}, Hong-Bo {Hu}, and Xiao-Jun {Bi}.
\newblock {Fermi Large Area Telescope Detection of Gamma-Ray Emission from the
  Direction of Supernova iPTF14hls}.
\newblock \emph{ApJl}, 854\penalty0 (2):\penalty0 L18, February 2018.

\bibitem[Zubko et~al.(1996)Zubko, Mennella, Colangeli, and
  Bussoletti]{zubko1996}
V~G Zubko, V.~Mennella, L.~Colangeli, and E~Bussoletti.
\newblock {Optical constants of cosmic carbon analogue grains - I. Simulation
  of clustering by a modified continuous distribution of ellipsoids}.
\newblock \emph{Monthly Notices of the Royal Astronomical Society},
  282\penalty0 (4):\penalty0 1321--1329, 1996.
\newblock ISSN 00358711.

\end{thebibliography}

\appendix*
\section{}

\begin{figure*}
\centering

\includegraphics[width=0.9\linewidth]{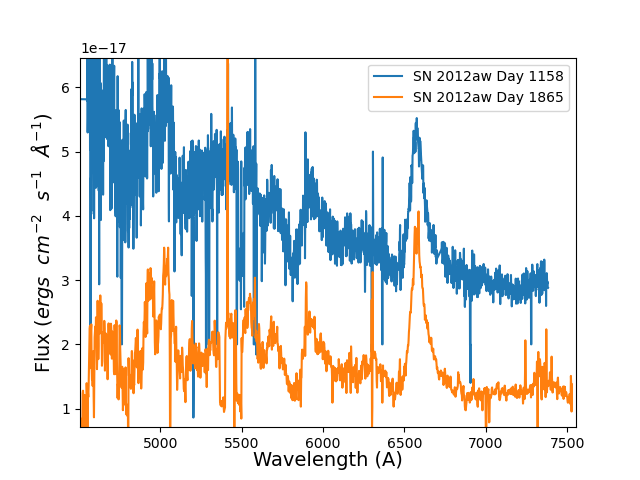}

\caption{The evolution of the optical spectrum of SN 2012aw. Information on the spectra is summarised in Section 3.}
\label{fig:12aw-fullspec}
\end{figure*}

\begin{figure*}
\centering

\includegraphics[width=0.9\linewidth]{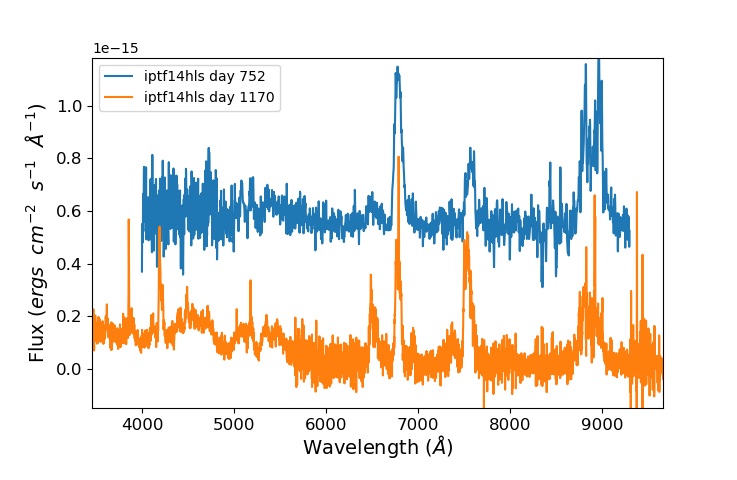}

\caption{The evolution of the optical spectrum of iPTF14hls. Information on the spectra is summarised in Section 3.}
\label{fig:iptf-fullspec}
\end{figure*}



\label{lastpage}
\end{document}